\documentclass[11pt,openany,openbib]{article}
\usepackage{amssymb}
\usepackage{amsmath}
\usepackage{epic}
\usepackage{eepic}
\usepackage{graphicx}
\DeclareSymbolFontAlphabet{\mathbb}{AMSb}
\usepackage{Vmargin}
\setmargins{1in}{1in}{6.5in}{9in}{0pt}{0pt}{12pt}{42pt}
\usepackage{times}
\usepackage{amsfonts}
\usepackage{eucal}
\usepackage{color}
\usepackage{pst-all}
\usepackage{latexsym}
\usepackage{amsbsy}

\newcommand{\qed}{\mbox{\rule{1.6mm}{4.3mm}}}

\usepackage{pseudocode}

\newtheorem{theo}{Theorem}
\newtheorem{lm}[theo]{Lemma}

\newtheorem{defn}[theo]{Definition}

\newtheorem{obs}[theo]{Observation}
\newtheorem{proof}{Proof}
\def\proof{{\bf Proof:~}}

\begin{document}

\title{To Broad-Match or Not to Broad-Match : An Auctioneer's Dilemma ?\footnote{A preliminary 
version of this paper was presented at Fourth Workshop on Ad Auctions.}} 

\author{Sudhir Kumar Singh\footnote{Department of Electrical Engineering, University of California, Los Angeles, CA 90095, ~~
Email:suds@ee.ucla.edu} ~ \footnote{Financially supported
by NetSeer Inc., Los Angeles, during course of this work.} \\
\and
Vwani Roychowdhury\footnote{Department of Electrical Engineering, University of California, Los Angeles, CA 90095, ~~
Email:vwani@ee.ucla.edu}
}

\maketitle

\begin{abstract}
We initiate the study of an interesting aspect of sponsored search advertising, namely the
consequences of {\it broad match-} a feature where an ad of an advertiser can be mapped to
a broader range of relevant queries, and not necessarily to the particular keyword(s)
that ad is associated with. In spite of its unanimously believed importance, this aspect
has not been formally studied yet, perhaps because of the inherent difficulty involved in
formulating a tractable framework that can yield meaningful conclusions. In this paper, we
provide a natural and reasonable framework that allows us to make definite statements
about the economic outcomes of broad match. 

Starting with a very natural setting for strategies available to the advertisers, and via
a careful look through the algorithmic lens,  we first propose  
solution concepts for the game originating from
the strategic behavior of advertisers as they try to optimize their budget allocation
across various keywords. 

Next, we consider two broad match scenarios based on
factors such as information asymmetry between advertisers and the auctioneer (i.e. the
search engine company), and the extent of auctioneer's control on the budget splitting. In
the first scenario,  the advertisers have the full information about broad match and
relevant parameters, and can reapportion their own budgets to utilize the extra information;
in particular, the auctioneer has no direct control over budget splitting. 
We show that, the same broad match may lead to
different equilibria, one leading to a \emph{revenue improvement}, whereas another to a
\emph{revenue loss}. This leaves the auctioneer in a \emph{dilemma} - whether to
broad-match or not, and consequently leaves him with a computational problem of predicting
which broad matches will provably lead to revenue improvement. This motivates us to
consider another broad match scenario, where the advertisers have information \textit{only}
about the current scenario, i.e., \emph{without} any broad-match, and the allocation of
the budgets unspent in the current scenario is in the control of the auctioneer. Perhaps
not surprisingly, we observe that if the quality of broad match is good, the auctioneer
can \textit{always} improve his revenue by judiciously using broad match. Thus, information
seems to be a double-edged sword for the auctioneer. Further, we also discuss the effect
of both broad match scenarios on social welfare.
\end{abstract}

\section{Introduction and Motivation}
\label{intro}
The importance of understanding the various aspects of 
sponsored search advertising(SSA) is now well known.
Indeed, this advertising framework has been studied extensively in recent years 
from algorithmic\cite{MSVV05, FMPS07, MPS07, RW06}, 
game-theoretic\cite{ EOS05, Varian06, AGM06, Lahaie06, CDE+07, EO07, LP07},
learning-theoretic\cite{GP07, WVLL07, SRBR08} perspectives, as well as, 
from the viewpoint of emerging diversification in the internet economy\cite{SRGR07, SRGR08}.  
Specifically,  among others, these studies include important aspects such as 
(i) the design of mechanisms for optimizing the revenue of 
the auctioneer, (ii) budget optimization problem of an advertiser, 
(iii) analyzing the bidding behavior of advertisers in the auction of a keyword query, 
(iv) learning the Click-Through-Rates and (v) the role of for-profit mediators.
In the present paper, our goal is to initiate the study of yet another interesting aspect of SSA,
namely the consequences of {\it broad match}. 
Despite its unanimously believed importance, to the best of our knowledge, 
this aspect has not been formally studied yet,
probably because of the hardness in employing a proper framework for such an study.
Our main aim in this paper is to attempt to provide such a framework.
In the rest part of this section, we start by an informal 
introduction to {\it broad match}, then while establishing the need of a framework to study {\it broad match}
we provide a glimpse of the present work.           

\subsection{What is Broad Match?}
\label{whatisbm}
In SSA format, each advertiser has a set of keywords relevant to her products and
a daily budget that she wants to spend on these keywords. Further, for each of these keywords 
she has a true value associated with it that she derives when a user 
clicks on her ad corresponding to that keyword, and based on this true value she 
reports her bid to the search engine company (i.e. the auctioneer) 
to indicate the maximum amount she is willing to pay for a click.   
When a user queries for a keyword, the auctioneer runs an auction among
all the advertisers interested in that keyword whose budget is not over yet.
The advertisers winning in this auction are allocated an ad slot each, determined 
according to the auction's allocation rule
and an advertiser is charged, whenever the user clicks on her ad,
an amount determined according to the auction's payment rule. 
In this way, for each of the advertisers,
each of her ads is matched to queries of the particular keyword(s) that ad is associated with. 

{\em Broad Match} is a feature where an ad of an advertiser can be mapped 
to a broader range of relevant queries, and not necessarily to the particular keyword(s) that ad is associated with.
Such relevant queries could be possible variations of the associated keyword or could 
even correspond to a completely different keyword which is {\it conceptually related} to the associated keyword.    
For example, the variations of keyword ``Scuba'' could include ``Scuba diving'',
``Scuba gear'', ``Scuba shops in Los Angeles'' etc 
and conceptually related keywords could include
``Snorkeling'', ``under water photography'' etc.
Similarly, for the keyword ``internet advertising'', the variations could include ``banner advertising'',
``PPC advertising'', ``advertising on the internet'', ``keyword advertising'', ``online advertising'' etc, and 
conceptually related keywords could include ``adword'', ``adsense'' etc;  
for the keyword ``horse race'', the variations could include ``horse racing tickets'',
``horse race betting'', ``online horse racing'' etc, and a conceptually related keyword could be ``thoroughbred''.
     
\subsection{The Need of a Framework to Study Broad Match }
\label{needframe}
To study the effect of incorporating {\it broad match} on macroscopic 
quantities such as revenue of the auctioneer, social value etc, compared 
to the scenario without broad match, we must consider the interaction among 
various keywords. We must take into account the changes in the bidding behavior 
of advertisers for specific keyword queries, as well as, the effect of 
changes in their budget allocation across various keywords. As we mentioned earlier,
the incentive properties of query specific keyword auction is well analyzed in articles 
such as\cite{ EOS05, Varian06, AGM06, Lahaie06, CDE+07, EO07, LP07}. 
Further, the budget optimization problem 
of an advertiser, that is to spend the budget across various keywords in 
an optimal manner given quantities such as keyword specific cost-per-clicks, expected number of clicks, and payoffs 
as a function of bid, has also been studied under various models\cite{FMPS07, MPS07, RW06}. 
However, the incentive constraints originating from such budget optimizing strategic 
behavior of advertisers has not been formally studied yet.
In particular, there is no proposed solution concept pertinent for predicting the stable 
behavior of advertisers in this game\cite{BCI+07}. On the other hand,
for the analysis of the effect of incorporating {\it broad match}, {\bf it becomes 
inevitable to have such a solution concept}, that is a notion of equilibrium
behavior, under which we can compare the quantities such as revenue of the auctioneer, social value etc,
for the two scenarios one with the broad match and the one without it at their respective 
equilibrium points. 

\subsection{Our Results}
\label{introresults}
Our first goal in this paper is to attempt to provide a reasonable 
solution concept for the game originating 
from the strategic behavior of advertisers as they try to optimize their budget allocation/splitting across 
various keywords. We realize that without some reasonable restrictions on the 
set of available strategies to the advertisers, it is a much harder task to achieve\cite{BCI+07}. 
To this end, we consider a very natural setting for available strategies- (i) first split/allocate the budget 
across various keywords and then (ii) play the keyword query specific bidding/auction game as long as you have budget left over 
for that keyword when that query arrives -thereby dividing the overall game in to two stages.
In the spirit of \cite{ EOS05, Varian06}
wherein the query specific keyword auction is modeled as a static one shot game of complete information despite 
its repeated nature in practice, we model the budget splitting too as a static one shot game of complete information
because if the budget splitting and bidding process ever stabilize, advertisers will be playing 
static best responses to their competitors' strategies. 

Now, with this one shot complete information game modeling, the most natural
solution concept to consider is {\em pure} Nash equilibrium. 
However, in our case, 
it seems to be a very strong notion of equilibrium behavior if we look through our
algorithmic glasses\footnote[1]{Efficient computability is an important modeling prerequisite 
for solution concepts. In the words of Kamal Jain, ``If your laptop cannot find it, neither can the market''\cite{papa07}.} 
because, as we argue in the paper, an advertiser's problem of choosing her best response is computationally hard.
Consequently, we first consider a weaker solution concept based
on \textit{local} Nash equilibrium. 
We show that there is 
a strongly polynomial time (polynomial in number of advertisers, keywords and ad slots and not 
the volume of queries or total daily budget)  algorithm to compute an advertiser's  \textit{locally}
best response. 
This notion of equilibrium, which we call {\it broad match equilibrium} (BME), 
is defined in terms of 
\textit{marginal payoff} (or \textit{bang-per-buck}) for various keywords corresponding to a 
budget splitting, and looks similar to the definition of \textit{user equilibrium/Waldrop equilibrium}
in routing and transportation science literature\cite{Roughgarden05}.
Further,  there is a strongly poly-time algorithm to compute an advertiser's \textit{approximate} best response as well.
Therefore, the \textit{approximate} Nash equilibrium ($\epsilon$-NE) is also reasonable in our setting.
Indeed, all the conclusions of our work hold true irrespective of which of these two solution concepts we adopt,
the \emph{BME} or the \emph{$\epsilon$-NE}. 

In the full information setting, under the solution concept of \emph{BME}
we obtain several observations by explicitly constructing examples.
In particular, even with this natural and reasonably restricted notion of stable behavior,
we observe that same \textit{broad match} might lead to different \emph{BME}s
where one \emph{BME} might lead to an improvement in the revenue of the auctioneer over the scenario without  
 \textit{broad match} while the other to a loss in revenue, even when the \textit{quality of broad-match}
is very good. This leaves the auctioneer in a dilemma about whether he should 
broad-match or not. If he could somehow predict which choice of broad match
lead to a revenue improvement for him and which not, he could potentially 
choose the ones leading to a revenue improvement.
\emph{Further, the same examples imply the same conclusions under the solution concept of $\epsilon$-NE}.
This brings 
forth one of the big questions left open in this paper, that is of efficiently computing a  \textit{BME} / \textit{$\epsilon$-NE}, if one exists, 
given a choice of \textit{broad match}.

Note that in the above scenario of broad match, 
the advertisers had all the relevant information about the 
broad match, and the control of budget splitting is in the hands of respective advertisers. 
We also explore another broad match scenario where the advertisers have information 
 \textit{only} about the scenario without broad-match, and they allow the auctioneer to 
spend their \textit{excess budgets} (i.e the budget unspent in the scenario without broad match),
in whatever way the auctioneer wants, in the hope of potential improvements in their payoffs.
Perhaps not surprisingly, we observe that if the quality of broad match is good,
the auctioneer can  \textit{always} improve his revenue. 
Thus, \textit{information seems to be a double-edged sword for the auctioneer}. 
Further, we also discuss the effect of 
both broad match scenarios on social welfare. 

\subsection{Organization of the Paper}
Rest of the paper is organized as follows. 
In Section \ref{apkeyauc}, we describe the formal setting of the query specific 
keyword auctions. 
In Section \ref{bmgframework}, we define {\em Broad Match Graph},
a weighted bipartite graph between the set of advertisers and the set of keywords, which serves as 
the basic back bone in terms of which we formulate all our notions and results. 
In Section \ref{twosc}, we present the defining characteristics of the two broad match scenarios studied in this paper, 
in terms of factors such as information asymmetry and the extent of auctioneer's control 
on the budget splitting. 
The Section \ref{adbm} is devoted to the study of {\it broad match} scenario in the full information setting
where the auctioneer has no direct control on the budget splitting. This is 
essentially the most contributing part of this paper wherein first we appropriately model 
the game originating 
from the strategic behavior of advertisers as they try to optimize their budget allocation/splitting across 
various keywords, then we propose appropriate solution concepts for this game, and finally
study the effect of broad match 
under this proposed framework. In Section \ref{acbm}, we study the other broad match scenario,
where advertisers do not have full information about the broad match being performed 
and the auctioneer partially controls the budget splitting. 
Finally, in Section \ref{concl}, we conclude with potential 
directions for future work motivated by the present paper.

\section{Keyword Auctions}
\label{apkeyauc}
There are $K$ slots to be allocated among  $N$ ($\geq K$)
bidders (i.e. the advertisers). A bidder $i$ has a true valuation $v_i$ (known only to the
bidder $i$) for the specific keyword and she bids $b_i$. The expected {\it click through
rate} (CTR) of an ad put by bidder $i$ when allocated slot $j$ has the form $CTR_{i,j}=\gamma_j e_i$
i.e. separable in to a position effect and an advertiser effect. $\gamma_j$'s can be
interpreted as the probability that an ad will be noticed when put in slot $j$ and it is
assumed that $\gamma_j > \gamma_{j+1}$ for all $1 \leq j \leq K$ and $\gamma_j = 0$ for $j > K$.
$e_i$ can be interpreted as the probability that an ad put by
bidder $i$ will be clicked on if noticed and is referred to as the {\it relevance} of bidder
$i$. The
payoff/utility of bidder $i$ when given slot $j$ at a price of $p$ per-click is given by
$e_i\gamma_j (v_i - p)$ and they are assumed to be rational agents trying to maximize
their payoffs. 

As of now, Google as well as Yahoo! use schemes closely modeled as RBR(rank by revenue)
with GSP(generalized second pricing). The bidders are ranked in the decreasing order of
$e_ib_i$ and the slots are allocated as per this ranks. For simplicity of notation, assume
that the $i$th bidder is the one allocated slot $i$ according to this ranking rule, then
$i$ is charged an amount equal to $\frac{e_{i+1} b_{i+1}}{e_i}$ per-click. This mechanism
has been extensively studied in recent years\cite{EOS05,Lahaie06,Varian06,LP07}.
The solution concept that is widely adopted to study this auction game is a refinement of
Nash equilibrium independently proposed by Varian\cite{Varian06} and Edelman et
al\cite{EOS05}. Under this refinement, the bidders have no incentive to change to another
positions even at the current price paid by the bidders currently at that position.
Edelmen et al \cite{EOS05} calls it  {\it locally envy-free equilibria} and argue that
such an equilibrium arises if agents are raising their bids to increase the payments of
those above them, a practice which is believed to be common in actual keyword auctions.
Varian\cite{Varian06} called it {\it symmetric Nash equilibria(SNE)} and provided some
empirical evidence that the Google bid data agrees well with the SNE bid profile. In
particular, an {\bf SNE} bid profile $b_i$'s satisfy
\begin{align}
\label{sne}
(\gamma_i - \gamma_{i+1}) v_{i+1} e_{i+1} + \gamma_{i+1} e_{i+2} b_{i+2}  \leq \gamma_{i} e_{i+1} b_{i+1} \nonumber \\
 \leq (\gamma_i - \gamma_{i+1}) v_{i} e_{i} + \gamma_{i+1} e_{i+2} b_{i+2}
\end{align}
for all $ i=1,2,\dots, N$.
Now, recall that in the RBR with GSP mechanism, the bidder $i$ pays an amount $\frac{ e_{i+1} b_{i+1} }{e_i}$ per-click,
therefore the expected payment $i$ makes per-impression is
$\gamma_i e_i \frac{ e_{i+1} b_{i+1} }{e_i} = \gamma_i e_{i+1} b_{i+1}$.
Thus the best {\bf SNE} bid profile for advertisers (worst for the auctioneer) is minimum bid profile
possible according to Equation \ref{sne} and is given by
\begin{equation}
 \gamma_{i} e_{i+1} b_{i+1} = \sum_{j=i}^K (\gamma_j - \gamma_{j+1}) v_{j+1} e_{j+1} \label{minsne0}
\end{equation}
and therefore, the revenue of the auctioneer at this minimum {\em SNE} is
\begin{align}
 \sum_{i=1}^K \gamma_{i} e_{i+1} b_{i+1} & = \sum_{i=1}^K \sum_{j=i}^K (\gamma_j - \gamma_{j+1}) v_{j+1} e_{j+1} \label{revminsne0}\\ \nonumber 
& = \sum_{i=1}^K  (\gamma_j - \gamma_{j+1}) j v_{j+1} e_{j+1}.
\end{align}

\section{Broad Match Graph}
\label{bmgframework}
In this section, we set up a basic backbone to study the dynamics of bidding across various {\it related} keywords,
in the sponsored search advertising, via a bipartite graph between the set of bidders (i.e. the advertisers)
and the set of keywords 
with the parameters $(N, M, K, S=(s_{i,j}), B=(B_i), V=(V_j))$ as follows. 

\begin{itemize}
\item {\bf Keywords:} $M$ is the number of keywords in consideration. We will denote the set of 
keywords $\{1,2,\dots,M\}$ by $\mathcal{M}$. Further, $V_j$ is the total (expected) volume of queries for
the keyword $j$ for a given period of time which we call a ``day''.  These may not be all the 
keywords a particular advertiser may be interested in bidding on but one of the sets of keywords which are 
related by some category/concept that the advertiser is interested in bidding on e.g. for selling the same product or service. 

\item {\bf Bidders/Advertisers:} $N$ is the total number of bidders interested in the above $M$ keywords.
We will denote the set of bidders $\{1,2,\dots,N\}$ by $\mathcal{N}$. Further, $B_i$ is the total budget 
of the bidder $i$ for a ``day'' that she wants to spend on these keywords. 

\item {\bf Slots:} $K$ is the maximum number of ad-slots available.
Let $L$ be the number of advertisers with sufficient budgets 
when a particular query of a keyword $j$ arrives then for that query of $j$, the slot-clickability 
(i.e. the position based CTRs) are defined to be $\gamma_l$ corresponding to a slot $l$ for $l \leq \min \{K,L\}$
and zero otherwise\footnote[2]{We assume that the clicks-through-rates(CTRs) are separable.}.   

\item {\bf Valuation Matrix:} Let $v_{i,j}$ be the true value of the bidder $i$ for the keyword $j$ and $e_{i,j}$ be her 
{\it relevance}\footnotemark[2] (quality score) for $j$. 
We call the $N \times M$ matrix $S$, with $(i,j)$th entries defined as 
$s_{i,j}= v_{i,j} e_{i,j}$, as the {\it valuation matrix}. For a keyword $j$ for which a bidder $i$ has no interest or is 
not allowed to bid, $s_{i,j}:= - \infty$. In the revenue/efficiency/payoffs calculations at SNE, it is enough to know
$s_{i,j}$ and not the $v_{i,j}$ and $e_{i,j}$ separately, therefore, in this paper we will often refer only to $s_{i,j}$ and 
not $v_{i,j}$ and $e_{i,j}$ individually.   

\item {\bf Broad Match Graph (BMG):}
Given instance parameters $(N, M, K, S, B, V)$, a bipartite graph $G=(\mathcal{N},\mathcal{M},\mathcal{E})$, with vertex
sets  $\mathcal{N}$ and $\mathcal{M}$ and edge set $\mathcal{E}=\{(i,j): i \in \mathcal{N}, j \in \mathcal{M}, s_{i,j} > 0\}$,
is constructed. 
Further, each edge $(i,j) \in \mathcal{E}$ is associated with a weight $s_{i,j}$. 
The weight of a node $i \in \mathcal{N}$ is $B_i$ and that of a node $ j \in \mathcal{M}$ is $V_j$. 
Furthermore, an $i \in \mathcal{N}$ will be called an {\it ad-node} and an $ j \in \mathcal{M}$ will be referred to as 
a {\it keyword-node}. 

\item {\bf Extension of a BMG:} 
A {\em BMG} $G^{'}=(\mathcal{N},\mathcal{M},\mathcal{E}^{'})$ for the instance $(N, M, K, S^{'}, B, V)$ is called an {\em extension}  
of a {\em BMG} $G=(\mathcal{N},\mathcal{M},\mathcal{E})$ for the instance $(N, M, K, S, B, V)$ 
if $\mathcal{E} \subset \mathcal{E}^{'}$ and $s_{i,j}^{'}=s_{i,j}$ for all
$(i,j) \in \mathcal{E}$. Interpretation is that the instance represented by an extension 
is more broader in the sense that an advertiser has a choice to be and could be matched to a larger set of keywords.  
Indeed, we will formally refer an {\em extension} $G^{'}=(\mathcal{N},\mathcal{M},\mathcal{E}^{'})$ to be a {\em broad-match} for its {\it base}  
 $G=(\mathcal{N},\mathcal{M},\mathcal{E})$. Note that, without loss of generality, this definition of extension (broad-match)
captures the possibility that new keywords are introduced. 
This is because we can always think that a new keyword to be included was already there as an isolated node and now we are creating only edges.
\end{itemize}

\section{Information Asymmetry, Budget Splitting and Two Broad Match Scenarios}
\label{twosc}
Depending on how much information the auctioneer (i.e. the search engine company)
provides to advertisers about various parameters,
such as CTRs, conversion rates, volume of queries etc, related to the keywords
and depending on who splits the daily budget across the various keywords, the auctioneer or the advertisers, we 
consider the following two scenarios for broad matching:  
\begin{itemize}
\item[I)] {\bf AdBM:} Advertisers Controlled Broad Match, extra information to advertisers,
 advertisers split their budgets. 

\item[II)] {\bf AcBM :} Auctioneer Controlled Broad Match, no extra information to advertisers, auctioneer splits the budget.
\end{itemize}

Formally, let
$G^{'}=(\mathcal{N},\mathcal{M},\mathcal{E}^{'})$ be the {\em broad-match} for
 $G=(\mathcal{N},\mathcal{M},\mathcal{E})$, then we characterize these two scenarios of broad-match 
based on the following factors(ref. {\bf Table \ref{twobmsc}}):

\begin{itemize}
\item {\bf {\it Information asymmetry:}} Of course, advertisers already have all the information about the {\em BMG} $G$.
That is, they all know their CTRs, conversion rates, volume of queries, who participates how long, spends and bids how much 
on which keyword, across various keywords they are bidding currently. It is the knowledge about 
these quantities for the part of the {\it extension} $G^{'}$ which is not in $G$ i.e. along the edges 
in $\mathcal{E}^{'} \setminus \mathcal{E}$ that advertisers may or may not be aware of depending on whether
the auctioneer provides this information to them or not\footnote[3]{Of course,
 to learn these relevance scores there might be a cost incurred by
the auctioneer in the short run, nevertheless if the broad-match quality is good, 
this cost will generally be minimal. }.
In {\bf AcBM} scenario, advertisers do not have this information.
At most what they know is that there is some broad-match the auctioneer is performing and they can effect the dynamics 
on the {\it extension} $G^{'}$ only passively through their bidding behavior on $G$\footnote[4]{As we know,
in the sponsored search advertising, advertisers effectively 
derive their valuations from the rate of conversion which might very well change by broad match and accordingly
bidders may adjust their true values (and consequently the bids) due to broad match. This might lead to another level of cost 
to the auctioneer due 
to uncertainty in performing broad-match. Nevertheless, 
if the broad-match quality is good, this cost will generally be minimal.}.  
In {\bf AdBM} scenario, there is no such asymmetry in information. 
Advertisers know (or are rather informed by the auctioneer)
all the information about $\mathcal{E}^{'} \setminus \mathcal{E}$. 

\item {\bf {\it Extent of auctioneer's control on the budget splitting:}} 
In {\bf AdBM} scenario, the auctioneer has no control over the budget splitting. 
It is upto the advertisers to decide which keywords they want to participate in and how much budget 
they want to spend on each of those keywords. Thus, the budget splitting of an advertiser, 
across various keywords she is connected to in $G^{'}$, is in her own control and
she can split her budget so as to maximize her total payoff across those keywords.
As discussed in section \ref{needframe}, this brings in another layer of
incentive constraints from advertisers besides choosing the bid values for individual queries.
In {\bf AcBM} scenario, the auctioneer does have some control over budget splitting. 
At one extreme, an advertiser might give the complete control to the auctioneer for
splitting her budget across various keywords letting the auctioneer decide how much 
should be spent on which keyword. In this case, since the control will be 
completely in the hands of the auctioneer it can perform the budget splitting 
so as to maximize his own total revenue without much concern to the welfare 
of advertisers. Therefore, advertisers would hardly give such a full control to
the auctioneer. However, it is reasonable that an advertiser allows the auctioneer 
to spend her {\it excess} budget i.e. the budget {\it unspent} in the case of $G$, in whatever way 
the auctioneer wants to spend it in $G^{'}$, in hope of an additional payoff. Note that there must be some 
advertisers with excess budget for the {\em AcBM} to make sense. If every advertiser 
is already spending its all budget then why to take all the pain of doing 
a broad-match\footnote[5]{In case, doing a broad-match encourages advertisers to increase their budgets, for the 
purpose of analysis, this increase 
can be considered as an excess budget.}.  
In our study of {\bf AcBM}  in Section \ref{acbm}, we consider this limited control setting. 
Another issue the auctioneer faces in {\bf AcBM} is that what bid profiles to use along the edges in 
$\mathcal{E}^{'} \setminus \mathcal{E}$. The auctioneer must perform these calculations         
in a manner to maintain equilibrium across the keywords meaning that the advertisers should not be indirectly compelled to
revise their true values along the edges in $G$\footnotemark[4]. This assumption is reasonable 
as at SNE the auctioneer can estimate the true values from the equilibrium bids on $G$ and then along with 
the new information gathered for $\mathcal{E}^{'} \setminus \mathcal{E}$, can do the proper SNE bids
calculations on the behalf of the advertisers.

\item {\bf { \it An advertiser's starting time for a keyword:}} In {\bf AdBM} scenario, an advertisers participates 
in all keywords she is interested in {\it starting with its first query} until her budget allocated for that 
keyword is spent or there are no more queries for that keyword\footnote[6]{In reality, Google/Yahoo! roughly allows the advertisers to specify 
which part of the day they want to spend most of their budgets, nevertheless this option does not give a finer control such as specifying 
a particular query number they can start with. Therefore, to simplify the incentive analysis we do not consider such option in this paper. Further, note that if the advertisers are given a chance to express their desire 
as to which part of the day they want to spend how much budget, and as long as the day is divided
in to few parts (say polynomially many in the size of the BMG), then this expressiveness can be easily 
captured in the present framework by replicating the role of relevant keyword nodes and dividing the total volume of
queries for that keyword node among these new nodes according to the size of the various parts of the day.}.   
In {\bf AcBM} scenario, however, since the auctioneer is controlling at least a part of the budget splitting, and unlike advertisers,
since it can easily 
track what happens at which query in an online manner, 
it can choose to bring in an advertiser's budget along an edge starting with any particular query of that 
keyword. For example, it can bring advertiser $i$ in the auction of keyword $j$ starting at say $1000$th query of $j$.      
\end{itemize}

\begin{table*}
\centering
\scalebox{0.8}{
\begin{tabular}{|c|c|c|}
\hline
 & {\bf AdBM} & {\bf AcBM} \\
\hline
{\em Information asymmetry}  & No & Yes  \\
  & (advertisers know all info on new edges) &  (advertisers don't know info on new edges)  \\
\hline
{\em Extent of auctioneer's control on the budget splitting} & No Control   & Limited Control  \\
 &  & (only for the excess budgets)  \\
\hline
{\em An advertiser's starting time for a keyword} & Very First Query & Any Query\\
 &  & (only for the part auctioneer controls)  \\
\hline
\end{tabular}
}
\caption{Two Broad-Match Scenarios}
\label{twobmsc}
\end{table*}

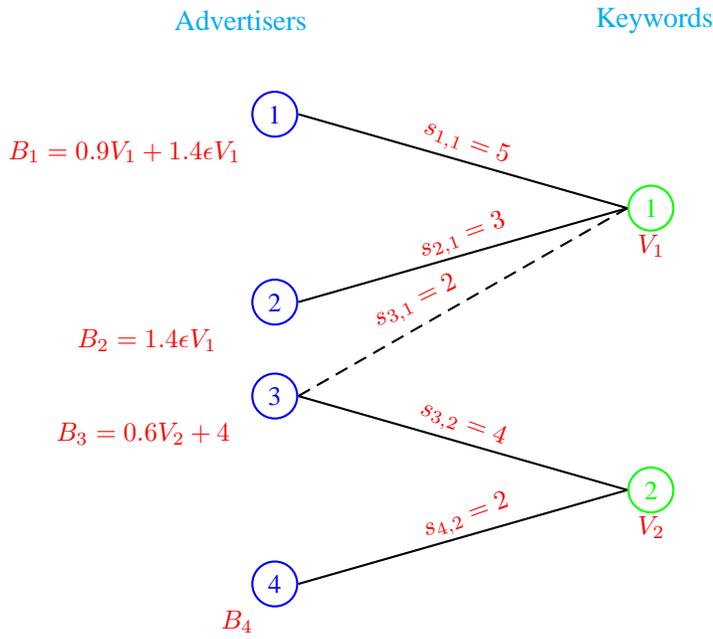
\begin{figure}[h] 
\begin{center} 
\psset{unit=1.25cm}
\begin{pspicture}(0, 0)(4, 6)
\psline(0.25,0)(3.75,1) 
\rput{20}(2,0.7){ \small \color{red} $s_{4,2}=2$}

\pscircle[linecolor=blue](0,0){0.25}
\rput(0,0){\small \color{blue} 4}
\rput(-0.4,-0.4){\small \color{red} $B_4$}

\pscircle[linecolor=green](4,1){0.25} 
\rput(4,1){\small \color{green} 2}
\rput(4,0.6){\small \color{red} $V_2$}

\psline(0.25,2)(3.75,1) 
\rput{-20}(2,1.7){\small \color{red} $s_{3,2}=4$}

\pscircle[linecolor=blue](0,2){0.25} 
\rput(0,2){\small \color{blue} 3}
\rput(-1.4,1.6){\small \color{red} $B_3=0.6 V_2+4$}

\psline(0.25,3)(3.75,4) 
\rput{20}(2,3.7){\small \color{red} $s_{2,1}=3$}

\pscircle[linecolor=blue](0,3){0.25} 
\rput(0,3){\small \color{blue} 2}
\rput(-1.4,2.6){ \small \color{red} $B_2=1.4 \epsilon V_1$}

\pscircle[linecolor=green](4,4){0.25} 
\rput(4,4){\small \color{green} 1}
\rput(4,3.6){\small \color{red} $V_1$}

\psline(0.25,5)(3.75,4) 
\rput{-20}(2,4.7){ \small \color{red} $s_{1,1}=5$}

\pscircle[linecolor=blue](0,5){0.25} 
\rput(0,5){\small \color{blue} 1}
\rput(-1.6,4.6){\small \color{red} $B_1 = 0.9V_1+1.4\epsilon V_1$}

\rput(-0.4, 6){ \color{cyan} Advertisers}
\rput(4, 6){ \color{cyan} Keywords}

\psline[linestyle=dashed](0.25,2)(3.75,4) 
\rput{26}(1.5,3){\small \color{red} $s_{3,1}=2$}

\end{pspicture}
\end{center} 
\caption{A {\em BMG} (without the edge (3,1) ) and an extension (with the edge (3,1)), $K=2, \gamma_1=1, \gamma_2=0.7$} 
\label{exfig02}
\end{figure}

Now, let us illustrate the above two scenarios via an example by 
considering the {\em BMG} given in Figure \ref{exfig02} (i.e. the graph without the edge $(3,1)$)
and its {\it extension} given in the same figure (i.e. the graph including the edge $(3,1)$).
Further, each query is sold via a {\em GSP} auction (ref. Section \ref{apkeyauc}), 
and the revenues are calculated at {\em SNE} (ref. Equations \ref{minsne0}, \ref{revminsne0} in Section \ref{apkeyauc}).  
In the base {\em BMG}, under {\it GSP} the advertiser $2$ pays zero amount for each query since there 
is no bidder ranked below her, therefore 
even with a very small budget she is able to participate in all the queries of keyword $1$. 
Thus, the total revenue extracted in the base {\em BMG} is 
$0.9V_1 + 0.6V_2$, $0.9V_1$ from keyword $1$ and  $0.6V_2$ from the keyword $2$. 
Now in the extension, there is a new edge $(3,1)$. In the {\em AdBM} scenario, 
since the advertiser $3$ has the control of splitting the budget and whether she 
wants to participate for keyword $1$, she participates along $(3,1)$ for all queries 
as she never needs to pay anything but certainly gets the second slot and positive payoffs 
for all the queries after the advertiser $2$ spends its all budget and drops out i.e.
for all the queries after the first $\epsilon V_1$th query.
Note that advertiser $2$ now pays a positive amount for each query and 
is forced to drop as its total budget gets spent. 
Therefore, the new revenue of the auctioneer 
is $0.9V_1+3.1V_1\left(\epsilon - \frac{0.3}{3.1}\right)+0.6V_2$ which is smaller than the revenue 
generated in the base {\em BMG} if $\epsilon < \frac{0.3}{3.1}$. 
However, in the {\em AcBM} scenario,
the control is in the hands of auctioneer and he can choose not to spend the (excess) budget of 
advertiser $3$ along $(3,1)$, thereby avoiding a potential revenue loss. Moreover, he has a finer 
control and can choose to bring $3$ along $(3,1)$ after some queries for $1$ has already 
arrived. In particular, if he brings $3$ along $(3,1)$ starting $(1-\epsilon)V_1+1$th query, the 
new revenue generated is $(1-\epsilon) 0.9V_1+2.3 \epsilon V_1+ 1.4 \epsilon V_1+0.6V_2$ which is more than 
the revenue generated in the base {\em BMG}. 

\section{Advertisers Controlled Broad Match (AdBM)}
\label{adbm}
In this section, we start out with our first goal, that is to provide a reasonable 
solution concept for the game originating 
from strategic behavior of advertisers as they try to optimize their budget allocation/splitting across 
various keywords. We realize that without some reasonable restrictions on the 
set of available strategies to the advertisers, it is a much harder task to achieve\cite{BCI+07}. 
To this end, we consider a very natural setting for available strategies- (i) first split/allocate the budget 
across various keywords and then (ii) play the keyword query specific bidding/auction game as long as you have budget left over 
for that keyword when that query arrives -thereby dividing the overall game in to two stages.

The second stage is exactly the query specific keyword auction and for that stage we can 
utilize the equilibrium behavior proposed in literature. Therefore, in the second stage, 
we are restricting the behavior of advertisers in that, given the  availability of budget to 
participate in the auction of a particular query, an advertiser acts rationally {\it but being ignorant}
about
and {\it disregarding }  
the fact that her bidding behavior might effect her decision in the splitting/allocation 
of her budget across various keywords in the first stage. For example, in the second stage, for the auction 
mechanisms currently used by Google and Yahoo! i.e. {\it Generalized Second Price} (GSP) Mechanism, 
we can adopt the solution concept of {\it Symmetric Nash Equilibrium(SNE)} proposed in \cite{ EOS05, Varian06}.
That is, once the budget is split, for each query all advertisers, with available budgets for the corresponding
keyword, bid according to a minimum {\em SNE} bid profile. {\it Note that, even for the same keyword, different queries 
may have different {\em SNE} bid profiles because the set of advertisers with available budgets may be
different when those queries arrive}. 

Now, for the first stage i.e. budget splitting, we do not put any restriction on how the
advertisers split their budget across various keywords. Each advertiser, {\it knowing the fact that 
all advertisers will behave according to SNE for keyword queries}, chooses 
her budget splitting so as to maximize her total payoff across all queries of all keywords 
for the day. Thus, with the restriction on the bidding behavior in second stage,
we are left only to analyze the game of splitting the budget across various keywords. 

In the spirit of \cite{ EOS05, Varian06}
wherein the query specific keyword auction is modeled as a static one shot game of complete information despite 
its repeated nature in practice, we model the budget splitting too as a static one shot game of complete information
because if the budget splitting and bidding process ever stabilize, advertisers will be playing 
static best responses to their competitors' strategies. Let us refer to this game as \textit{Broad-Match Game}.
In the following, we search for a reasonable notion of stable budget splitting
 i.e. a reasonable solution concept for the Broad-Match Game. 

\subsection{Advertisers' Best Response Problem and Search for an Appropriate Solution Concept for Broad-Match Game }
\label{adbrp}

Let the budget of advertiser $i$ decided for the keyword $j$ be $B_{i,j}$, then this budget allows her to participate 
for the first $V_{i,j}$ number of queries for some $V_{i,j} \leq V_j$. And given a number of queries $V_{i,j}$ for 
$j$, there will be a budget requirement from $i$, depending on the bidding interests of the other bidders, their valuations etc,
to participate in the first  $V_{i,j}$ queries of $j$. Thus, there is a one-to-one correspondence between the 
budget spent on a particular keyword by a bidder and the total number of queries of that keyword starting the first one, 
that the bidder participates in. Clearly, then the splitting of budget across various keywords
can be equivalently considered as deciding on how many queries 
to participate in, starting their first queries, 
for those keywords.

Now, with the one shot complete information game modeling, the most natural
solution concept to consider is {\em pure} Nash equilibrium which in our 
scenario (i.e. for the Broad-Match Game corresponding to the instance $(N, M, K, S, B, V)$ \&  
BMG $(\mathcal{N},\mathcal{M},\mathcal{E})$) can be defined 
as the matrix of query values $\{V_{i,j}, (i,j) \in \mathcal{E}\}$  such that for each $i$,
given that these values are fixed for all $ l \in \mathcal{N} - \{i\}$, no other values of $\{V_{i,j}\}$ 
gives a better total payoff to the bidder $i$. Equivalently,
it is the  matrix of values $\{B_{i,j}, (i,j) \in \mathcal{E}\}$ such that for each $i$,
given that the these values are fixed for all $ l \in \mathcal{N} - \{i\}$, no other values of $\{B_{i,j}\}$ 
gives a better total payoff to the bidder $i$.  

The solution concept of pure Nash equilibrium that we have proposed above requires 
that the players (i.e. the advertisers) are powerful enough to play the game rationally.
In particular, they should be able to compute the best responses to their competitors' strategies.
In the present case, for an advertiser $i$, it boils down to solving an optimization problem 
of finding the budget-splitting  across various keywords with maximum total payoff, given the budget- 
splitting of other advertisers. 
That is, given the values $\{B_{n,j}\}$ for all $n \in \mathcal{N}-\{i\}$ and for all $j \in \mathcal{M}$,
to compute $\{B_{i,j}\}$ yielding maximum payoff for the advertiser $i$.  
Of course, the details of {\em BMG} is known to every player.
To justify this solution concept on algorithmic 
grounds,
it should be essential that this optimization problem of advertiser $i$, henceforth referred to as
{\em Advertisers' Best Response Problem(AdBRP)},
should be solvable in time polynomial in $N, M, K$. Note that we are strictly asking the time complexity  
to be polynomial in $N, M, K$ and not in $V_j$'s and $B_i$'s. This is because $V_j$'s and $B_i$'s could in general 
be exponentially larger than 
 $N, M, K$.  In fact, in practice this seem to be the case as volume of queries could be much larger 
than the number of advertisers and keywords.  
Therefore, let us first formulate this optimization problem for advertiser $i$.  
Let $c(m,j,l)$ be the (expected) cost of advertiser $m$ for the $l$th query of the keyword $j$ if she participates in 
the query $l$ and all the previous queries of the keyword $j$. Similarly, 
 $u(m,j,l)$ is the (expected) payoff of advertiser $m$ for the $l$th query of the keyword $j$ if she participates in 
the query $l$ and all the previous queries of the keyword $j$. Also define the {\em marginal payoff} or
{\em bang-per-buck} for advertiser $m$ for the $l$th query of keyword $j$ as $\pi(m,j,l) = \frac{u(m,j,l)}{c(m,j,l)}$. 
To be able to compute
her best response, an advertiser must be able to efficiently compute these values first.
For each $(i,j) \in \mathcal{E}$ there are $O(V_j)$ such values and at first glance it seems that the advertiser might take 
a time polynomial in $V_j$, i.e. not in strongly polynomial time, to compute these values. However, if we observe 
carefully there is a lot of redundancy in these values as noted in the 
following lemma. 
The idea is that the cost and the payoff of an advertiser for a query of a keyword depends only
on the set of advertisers participating in that query and the valuations their of.
Therefore, as long as 
this set is the same these quantities will remain the same.
\begin{lm}
\label{par1}
For each $i \in \mathcal{N}$, given the {\em BMG} $G=(\mathcal{N}, \mathcal{M}, \mathcal{E})$
and the budget splitting of all advertisers in $\mathcal{N} - \{i\}$
(i.e. the $B_{n,j}$ values for all $n \in \mathcal{N}, j \in \mathcal{M}$),  
there exist non-negative integers $\Lambda_j$, $z_{j,0}, z_{j,1}, z_{j,2}, \dots, z_{j,\Lambda_j}$
for all $j \in \{j \in \mathcal{M}:(i,j) \in \mathcal{E}\}$ such that $\Lambda_j = O(N)$ and 
 for all $0 \leq \lambda \leq \Lambda_j-1$, 
\begin{equation*}
c(i,j,l) = C(i,j,\lambda) , u(i,j,l) = U(i,j,\lambda), \pi(i,j,l) = \Pi(i,j,\lambda) ~~~
\forall
 z_{j,\lambda} < l \leq z_{j,\lambda+1},   
\end{equation*} 
where $C(i,j,\lambda) := c(i,j,z_{j,\lambda}+1) , U(i,j,\lambda):=u(i,j,z_{j,\lambda}+1), \Pi(i,j,\lambda):=\pi(i,j,z_{j,\lambda}+1)$.
Moreover, all these $\Lambda, z, c, u, \pi$ values can be computed together in $O(M N^2 K)$ time.   
\end{lm}  

\proof 
The proof follows from  Algorithm \ref{algoqp} described in the following. 
The idea in the Algorithm \ref{algoqp} is that the cost and the payoff of an advertiser for a query of a keyword depends only
on the set of advertisers participating in that query and the valuations their of.
Therefore, as long as 
this set is the same these quantities will remain the same. Further, there can be at most $N$ choices 
of this set. For the very first query, this set consists of all the advertisers with 
positive budget for that keyword along with the advertiser $i$ (recall that the budget 
splittings for other advertisers are given).
This set changes when advertisers drop out as their budgets get spent and they no longer have enough amount to buy the next query. 
This change can occur 
at most $N-1$ times. The Algorithm \ref{algoqp} tracks exactly when the advertisers drop out and 
compute the $c,u,\pi$ values accordingly. \\

The Algorithm \ref{algoqp} runs in time $O(N^2 K)$. Initialization requires $O(N)$ and the $\WHILE$ loop requires $O(NK)$ 
in each iteration. There are at most $O(N)$ iteration of the $\WHILE$ loop because for every two 
iteration at least one advertiser drops out.
There is a $NlogN$ term, subsumed by $N^2$, that comes from the need to sort the valuation $s_{i,j}$'s before computing the SNE bid profile. \\

\begin{pseudocode}[Ovalbox]{Query Partition}{G,j, B_{n,j} \forall n \in \mathcal{N}-\{i\}}
\label{algoqp}
 A \GETS \{ n \in \mathcal{N} - \{i\}: B_{n,j} > 0\} \\
\lambda \GETS 0 \\
z_{j,0} \GETS 0 \\
\\
\WHILE A \neq \phi \DO
\BEGIN
Compute ~~
C(n,j,\lambda)  ~~  \forall n \in A \cup \{i\} \\
\textrm{ and } U(i,j,\lambda) \\
\\
\Pi(i,j,\lambda) \GETS \frac{U(i,j,\lambda)}{C(i,j,\lambda)}\\
\\
\IF C(i,j,\lambda) =0 \THEN 
\Pi(i,j,\lambda) \GETS \infty\\
\\
y \GETS \min_{n \in A} \{ \lfloor \frac{B_{n,j}}{C(n,j,\lambda)} \rfloor \} \\
\\
\IF y > 0 \THEN 
\BEGIN
A_1 \GETS A - \{ n \in A:  \frac{B_{n,j}}{C(n,j,\lambda)} = y\} \\
B_{n,j} \GETS B_{n,j} - y C(n,j,\lambda)  ~~ \forall n \in A \\
z_{j,\lambda+1} \GETS \min \{z_{j,\lambda} + y, V_j\} \\
A \GETS A_1 \\
\lambda \GETS \lambda +1 \\
\\
\IF z_{j,\lambda}= V_j \THEN 
\EXIT \\
\COMMENT{ Exiting the $\WHILE$ LOOP}
\END
\ELSE 
\\
A \GETS \{ n \in A : \lfloor \frac{B_{n,j}}{C(n,j,\lambda)} \rfloor \neq 0 \} \\
\END
\\
\\
\IF A = \phi \THEN 
\BEGIN
C(i,j,\lambda) \GETS 0  \\
U(i,j,\lambda) \GETS \gamma_1 s_{i,j} \\
\Pi(i,j,\lambda) \GETS \infty \\
\lambda \GETS \lambda +1 \\
z_{j,\lambda} \GETS V_j \\
\END
\\
\\
\Lambda_j \GETS \lambda\\
\\
\OUTPUT{\Lambda_j, \{z_{j,\lambda}\}, \{C(i,j,\lambda)\}, \{U(i,j,\lambda)\}, \{\Pi(i,j,\lambda)\} }
\end{pseudocode}

\qed.

Now, using the above lemma we are ready to formulate the {\em AdBRP} of advertiser $i$.
Let us define, 
{\small 
\begin{eqnarray}
\tilde{U}(i,j,l) & = & ( l - z_{j,\lambda-1}) U(i,j,\lambda -1)  +  \sum_{m=1}^{\lambda-1} (z_{j,m} - z_{j,m-1}) U(i,j,m-1) 
  ~~ \textrm{if  } z_{j,\lambda-1} < l \leq z_{j,\lambda}. \label{utfun}\\
\nonumber \\
\tilde{C}(i,j,l) & =  & ( l - z_{j, \lambda-1}) C(i,j,\lambda -1)
 + \sum_{m=1}^{\lambda-1} (z_{j,m} - z_{j,m-1}) C(i,j,m-1) 
  ~~ \textrm{if  } z_{j,\lambda-1} < l \leq z_{j,\lambda}.  \label{ctfun} 
\end{eqnarray}
}
then the  {\em AdBRP}  is the following optimization problem in non-negative integer variables $x_{i,j}$'s.
\begin{eqnarray}
\textrm{ {\em Max}   }  & &   \sum_{(i,j) \in \mathcal{E}} \tilde{U}(i,j,x_{i,j}) \nonumber \\
& & \nonumber \\
 \textrm{ {\em s.t.}   }  & &  \sum_{(i,j) \in \mathcal{E}} \tilde{C}(i,j,x_{i,j}) \leq B_i \label{budcons} \\
& & \nonumber \\
& & 0 \leq  x_{i,j} \leq V_j  ~~~ \forall ~ (i,j) \in \mathcal{E}\nonumber \\   
& & \nonumber \\
 & & x_{i,j} \in \mathcal{Z}  ~~~ \forall ~ (i,j) \in \mathcal{E}\nonumber 
\end{eqnarray}

It is not hard to see that being a variant of {\it Knapsack Problem} 
(decision version of) {\em AdBRP} is {\it NP-hard}.  
Its NP hardness follows from simple restrictions that makes it equivalent to  0-1 Knapsack Problem 
or Integer Knapsack Problem(IKP) respectively. 
First, let us restrict all relevant quantities such as budget, utilities and costs to be integer values. 
Now, in AdBRP, choosing $V_j=1$ for all $ j \in \mathcal{M}$ makes it equivalent to the 
 0-1 Knapsack Problem. Also, in AdBRP,
choosing $\Lambda_j =1$ for all $ j \in \mathcal{M}$ (i.e. a single partition for each keyword)
makes it equivalent to the Integer Knapsack Problem. \\

Now being NP-hard, {\em AdBRP} is unlikely to have an efficient
algorithm
and thus the solution concept based on pure Nash equilibrium 
does not seem to be  reasonable
and we should consider weaker notions.
One such reasonable solution concept 
could be that based on local Nash equilibrium, where the advertisers are 
not required to be so sophisticated and can deviate only locally i.e. by small amounts from the 
their current strategies. We show that an advertiser's \textit{locally best response}
can be computed via a greedy algorithm in strongly polynomial time.
Further, this equilibrium notion is similar to the \textit{user equilibrium/Waldrop equilibrium}
in routing and transportation science literature\cite{Roughgarden05}. 
Another solution concept that we explore is motivated by the fact that being a variant of \emph{IKP} and given that 
\emph{IKP} can be approximated well, it may be possible to efficiently compute a pretty good approximation 
of the advertiser's best response, and therefore an approximate Nash equilibrium may also make sense. 
In the following sub-sections we investigate these two solution concepts.

\subsection{Broad Match Equilibrium(BME)}
\label{bme}
Based on our discussion in section \ref{adbrp}, let us first formally define the equilibrium notion based on local 
Nash equilibrium and let us refer to it as \emph{Broad Match Equilibrium(BME)}. 
\begin{defn}
Given a {\em BMG} $G=(\mathcal{N},\mathcal{M},\mathcal{E}$), a {\em BME} for $G$ is defined 
as the matrix of query values $\{V_{i,j}\}$ and the budget splitting values $\{B_{i,j}\}$ iff they satisfy 
the following conditions.

\begin{itemize}
\item[E1)] For all $i$, if $(i,j) \in \mathcal{E}$ with $V_{i,j} > 0$, and $(i,l) \in \mathcal{E}$
 which is not query-saturated meaning $V_{i,j} < V_j$ then 
$MP_{i,j}^{-} := \frac{u(i,j,V_{i,j})}{c(i,j, V_{i,j})} \geq MP_{i,l}^{+} := \frac{u(i,l, V_{i,l}+1)}{c(i,l,V_{i,l}+1)}$ 
i.e. the advertiser $i$ does not have an incentive to 
deviate locally from keyword $j$ to keyword $l$.

\item[E2)] For all $i$, $i$ spends her total budget $B_i$ (on some keyword or the other) unless 
each $(i,j) \in \mathcal{E}$ is either query-saturated (i.e. $V_{i,j}=V_j$) or
budget-saturated meaning that the left over budget of advertiser $i$ for keyword $j$ is insufficient to buy the next query of $j$. 
\end{itemize}
\end{defn}       
Now given the $\{V_{n,j}\}$ and $\{B_{n,j}\}$ values for all $n \in \mathcal{N}-\{i\}$, 
the locally best response problem ( \emph{local AdBRP})
for $i$ is to compute a set of $\{V_{i,j}\}$ and $\{B_{i,j}\}$ values such that they 
satisfy all conditions in the above definition of {\em BME}.
We show in the following theorem that the \emph{local AdBRP} can be computed efficiently. 
Thus, {\it BME} is indeed a reasonable solution concept for the Broad-Match Game.

The idea in the proof of Theorem \ref{gsplittheorem} is to 
first partition the queries across various keywords via  Lemma \ref{par1} so that the payoff, 
cost, and marginal payoff of $i$ is the same for all queries in a given partition. 
Then, in a {\em GREEDY ALLOCATION PHASE}, to greedily distribute the budgets across various keywords moving from one 
partition to another. Finally, in a {\em GREEDY READJUSTMENT PHASE}, if there is an edge $(i,l)$ which is
not stable (i.e. the advertiser $i$ could profitably deviate locally to this edge from another edge(s)),  
a reverse greedy approach would take budgets from edge with minimum marginal payoff and put it to $(i,l)$ 
until $(i,l)$ becomes stable,  again via moving from partition to partition.
Since we always move from partition to partition 
and a partition is visited at most once in each of the above two phases, 
and there are at most $O(N)$ partitions per keyword,
this algorithm is efficient.  

\begin{theo}
\label{gsplittheorem}
There is a strongly polynomial time algorithm for local AdBRP. 
\end{theo}
\proof 
The proof follows from the Algorithm \ref{algogbs} provided in the following. 
The Algorithm \ref{algogbs}  first partitions the queries across various keywords so that the payoff, 
cost, and marginal payoff of $i$ is the same for all queries in a given partition. 
This is computed via Algorithm \ref{algoqp} which takes $O(MN^2K)$ time.
Initialization takes $O(M)$ time.   
In the {\em GREEDY ALLOCATION PHASE}, the algorithm greedily distributes the budgets across various keywords moving from one 
partition to another. Each iteration of the $\WHILE$ loop in this phase takes $O(M)$ time and since there are $O(MN)$ partitions
the total time taken for this loop is  $O(M^2N)$. {\em GREEDY READJUSTMENT PHASE} first checks if there is any edge $(i,l)$ which is
not stable i.e. the advertiser $i$ could profitably deviate locally to this edge from another edge(s). It is not hard to see 
that at the end of {\em GREEDY ALLOCATION PHASE}, there can be at most one such edge
(ref. Lemma \ref{oneuedge}). If there is such an edge, this phase
adjusts the budget to make this edge stable without making any other edge unstable. This is achieved via 
a reverse greedy approach taking budgets from edge with minimum marginal payoff and putting to $(i,l)$ 
until $(i,l)$ becomes stable,  again via moving from partition to partition.
Thus, \emph{Algorithm \ref{algogbs} correctly computes a locally best response for advertiser $i$.} 
 The $\WHILE$ loop in this readjustment phase also terminates in 
$O(MN)$ iterations and therefore this phase takes total of $O(M^2N)$ time. Hence, the running time of Algorithm \ref{algogbs} is 
$O(MN^2K+M^2N)$ i.e. strongly polynomial time.

\begin{pseudocode}{Greedy Budget Splitting}{G, B_{n,j} \forall n \in \mathcal{N}-\{i\}, j \in \mathcal{M}}
\label{algogbs}
J_i = \{ j \in \mathcal{M}: (i,j) \in \mathcal{E} \} \\
\\
\COMMENT{Computing $\Lambda, z, C, U, \Pi$ values for all relevant $j$'s} \\
\FOREACH j \in J_i \DO \CALL{Query Partition}{G,j, B_{n,j} \forall n \in \mathcal{N}-\{i\}}\\
\\
\COMMENT{Initialization} 
\\
\lambda_j \GETS 0 ~~ \forall j \in J_i \\
y_j \GETS z_{j,1} ~~ \forall j \in J_i \\
C_{i,j} \GETS 0, B_{i,j} \GETS 0, V_{i,j} \GETS 0 ~~ \forall j \in J_i \\
C_i\GETS 0\\
\\
\COMMENT{{\em GREEDY ALLOCATION PHASE}}
\\
\WHILE J_i \neq \phi \DO
\BEGIN
L \GETS arg \max_{j \in J_i} \Pi(i,j,\lambda_j) \\ 
\textrm{ if there are more than one such index take the minimum one} \\
\\
\IF C_i + y_L C(i,L, \lambda_L) > B_i \THEN 
\BEGIN
y \GETS \lfloor \frac{B_i - C_i}{C(i,L,\lambda_L)}\rfloor \\
V_{i,L} \GETS V_{i,L} + y \\
B_{i,L} \GETS B_{i,L} + (B_i -C_i) \\
C_{i,L} \GETS C_{i,L} + y C(i,L, \lambda_L) \\ 
C_i \GETS C_i +  y C(i,L, \lambda_L) \\
\EXIT \\
\COMMENT{ Exiting the $\WHILE$ LOOP}
\END
\ELSE 
\BEGIN
B_{i,L} \GETS B_{i,L} + y_L C(i,L, \lambda_L) \\
C_{i,L} \GETS C_{i,L} + y_L C(i,L, \lambda_L) \\
C_i \GETS C_i + y_L C(i,L, \lambda_L) \\
V_{i,L} \GETS V_{i,L} + y_L \\
\\
\IF \lambda_L = \Lambda_j -1 \\
\COMMENT{ i.e. no more queries for $L$ is left}
\THEN 
\BEGIN
J_i \GETS J_i - \{L\} \\
\lambda_L \GETS \lambda_L+1 \\
\END
\ELSE 
\BEGIN
y_L \GETS z_{L,\lambda_L+1} - z_{L,\lambda_L}\\
\lambda_L \GETS \lambda_L+1 \\
\END
\END
\END
\end{pseudocode}

\begin{pseudocode}[display]{}{}
\COMMENT{{\em GREEDY READJUSTMENT PHASE}} 
\\
J_i \GETS \{ j: (i,j) \in \mathcal{E} \AND ~ V_{i,j} > 0 \} \\
\\
l \GETS \{ l \in \mathcal{M}: (i,l) \in \mathcal{E} \AND ~ \exists j \in J_i - \{l\} ~ s.t. ~ MP_{i,l}^{+} > MP_{i,j}^{-} \} \\
\COMMENT{there can be at most one such $l$ (ref. Lemma \ref{oneuedge})}\\
\\
\IF \textrm{ such an $l$ does not exist } \THEN \RETURN{\{V_{i,j}\},\{B_{i,j}\}} \\
\COMMENT{Readjustment not required}
\\
\\
\WHILE \lambda_l < \Lambda_l \AND MP_{i,l}^{+} > \min_{j \in J_i -\{l\}} MP_{i,j}^{-} 
\\
\DO
\BEGIN
j \GETS arg \min_{j \in J_i-\{l\}} MP_{i,j}^{-} \\
\\
y \GETS \left\lfloor \frac{(V_{i,j} - z_{j,\lambda_j -1}) C(i,j,\lambda_j -1) + (B_{i,j} -C_{i,j}) + (B_{i,l}-C_{i,l})}{C(i,l,\lambda_l)}\right\rfloor \\
\\ 
\IF y < z_{l,\lambda_l+1} -V_{i,l} \THEN 
\BEGIN
B_{i,l} \GETS B_{i,l} + (V_{i,j} -z_{j,\lambda_j-1}) C(i,j,\lambda_j -1)\\
 + (B_{i,j} -C_{i,j}) \\
C_{i,j} \GETS C_{i,j} - (V_{i,j} -z_{j,\lambda_j-1}) C(i,j,\lambda_j -1) \\
B_{i,j} \GETS C_{i,j}\\
C_{i,l} \GETS C_{i,l} + y C(i,l,\lambda_l) \\
\lambda_j \GETS \lambda_j-1 \\
V_{i,j} \GETS z_{j,\lambda_j} \\
\\
\IF V_{i,j} =0 \THEN J_i \GETS J_i -\{j\} \\
\\
V_{i,l} \GETS V_{i,l}+y \\
\END
\\
\ELSE 
\BEGIN
\tilde{y} \GETS \left\lceil \frac{(z_{l,\lambda_l+1}-V_{i,l}) C(i,l,\lambda_l) - (B_{i,j}-C_{i,j}) - (B_{i,l}-C_{i,l})}{C(i,j,\lambda_j-1)}\right\rceil \\
\\
C_{i,j} \GETS C_{i,j} - \tilde{y} C(i,j,\lambda_j-1) \\
B_{i,j} \GETS B_{i,j} + \left[ (z_{l,\lambda_l+1} - V_{i,l}) C(i,l,\lambda_l) - (B_{i,l} -C_{i,l}) \right] \\
C_{i,l}  \GETS C_{i,l} + (z_{l,\lambda_l+1} - V_{i,l}) C(i,l,\lambda_l) \\
B_{i,l}  \GETS C_{i,l}\\
\\
\IF \tilde{y} = V_{i,j} - z_{j,\lambda_j -1} \THEN
\BEGIN
 \lambda_j \GETS \lambda_j-1 \\
V_{i,j} \GETS z_{j,\lambda_j} \\
\IF V_{i,j} =0 \THEN  J_i \leftarrow J_i -\{j\} \\ 
\END
\\
\ELSE 
 V_{i,j} \GETS V_{i,j} - \tilde{y}\\ 
\\
\lambda_l \GETS \lambda_l+1 \\
V_{i,l} \GETS z_{l,\lambda_l} \\
\\
\END
\END
\\
\\
\OUTPUT{\{V_{i,j}\},\{B_{i,j}\}}
\end{pseudocode}

\qed.

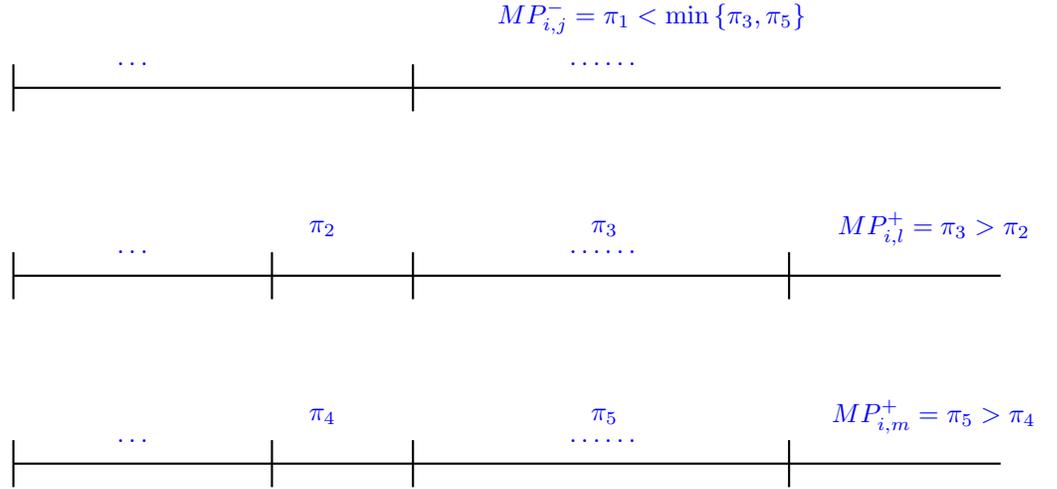
\begin{figure}[h] 
\begin{center} 
\psset{unit=1.25cm}
\begin{pspicture}(-1, -1)(4, 4)

\psline(-2.25,3)(8.25,3) 
\psline(-2.25,2.75)(-2.25,3.25)
\psline(2,2.75)(2,3.25) 
\rput(-1,3.25){ \small \color{blue} $\dots$}
\rput(4.5,3.75){ \small \color{blue} $MP_{i,j}^{-}=\pi_1 < \min \left\{ \pi_3, \pi_5 \right\}$}
\rput(4,3.25){ \small \color{blue} $\dots \dots $}

\psline(-2.25,1)(8.25,1) 
\psline(-2.25,0.75)(-2.25,1.25)
\psline(0.5,0.75)(0.5,1.25) 
\psline(2,0.75)(2,1.25) 
\psline(6,0.75)(6,1.25) 
\rput(-1,1.25){ \small \color{blue} $\dots$}
\rput(1,1.5){ \small \color{blue} $\pi_2$}
\rput(4,1.5){ \small \color{blue} $\pi_3$}
\rput(4,1.25){ \small \color{blue} $\dots \dots $}
\rput(7.5,1.5){ \small \color{blue} $MP_{i,l}^{+}=\pi_3 > \pi_2$}

\psline(-2.25, -1)(8.25, -1) 
\psline(-2.25,-1.25)(-2.25,-0.75)
\psline(0.5,-1.25)(0.5,-0.75) 
\psline(2,-0.75)(2,-1.25) 
\psline(6,-0.75)(6,-1.25) 
\rput(-1,-0.75){ \small \color{blue} $\dots$}
\rput(1,-0.5){ \small \color{blue} $\pi_4$}
\rput(4,-0.5){ \small \color{blue} $\pi_5$}
\rput(4,-0.75){ \small \color{blue} $\dots \dots $}
\rput(7.5,-0.5){ \small \color{blue} $MP_{i,m}^{+}=\pi_5 > \pi_4$}

\end{pspicture}
\end{center} 
\caption{There can not be two unstable edges at the end of {\em GREEDY ALLOCATION PHASE} in Algorithm \ref{algogbs}. 
The $\pi$ values shown in the figure are the marginal payoffs of $i$ in the respective partitions. ``$\dots$''  shown between 
two partitions indicate 
that there could be several or no partitions between these partitions. } 
\label{unst1}
\end{figure}

\begin{lm}
\label{oneuedge}
At the end of {\em GREEDY ALLOCATION PHASE } of Algorithm \ref{algogbs}, there can be at most one unstable edge, that is 
\begin{align*}
|\left\{ l \in \mathcal{M}: (i,l) \in \mathcal{E} \AND ~ \exists j ~
s.t. (i,j) \in \mathcal{E}, V_{i,j} > 0 ~ \&  ~ MP_{i,l}^{+} > MP_{i,j}^{-} \right\}|
 < 1. 
\end{align*}
\end{lm}
\proof 
We will provide a proof by contradiction. 
If possible, let there be two unstable edges namely $(i,l)$ and $(i,m)$, thus there exist an edge $(i,j) \in \mathcal{E}$ with $V_{i,j} > 0$ such that 
$MP_{i,l}^{+} > MP_{i,j}^{-}$ and  $MP_{i,m}^{+} > MP_{i,j}^{-}$. Note that there could be many such choices of $j$, let us 
choose the one with the minimum 
value of $MP_{i,j}^{-}$. Thus, in the Figure \ref{unst1}, we are given that  
$\pi_1 < \min \left\{\pi_3, \pi_5 \right\}$.  
Since the greedy allocation phase fills up ( or selects) the partitions with higher values first, starting the markers 
on the first partition of all the keywords,
there must be partitions with values $\pi_2 \leq \pi_1$ and $\pi_4 \leq \pi_1$ as shown in Figure \ref{unst1},
otherwise all partitions of $(i,l)$ before and with the value $\pi_3$ must have been filled/selected before
the partition with value $\pi_1$ and similarly for all partitions of $(i,m)$ before and with the value $\pi_5$.
WLog let the partition with the value $\pi_2$ is filled before that with $\pi_4$. 
Therefore, $\pi_3 \leq \pi_4$ otherwise the open partition ( i.e. the one not yet completely filled ) 
with value $\pi_3$ would have been filled before $\pi_4$.
Further, by using $\pi_2 \leq \pi_1$ and $\pi_1 < \pi_3$ we get $\pi_2 < \pi_3$ and consequently  $\pi_2 < \pi_4$.
As before we can again argue that 
there exist a partition with value $\pi_6 < \pi_2$ but we must also have $\pi_3 \leq \pi_6$ otherwise 
the open partition with value $\pi_3$ would have been filled before $\pi_6$. But this implies 
$\pi_3 < \pi_2$ which is a contradiction.  \qed.

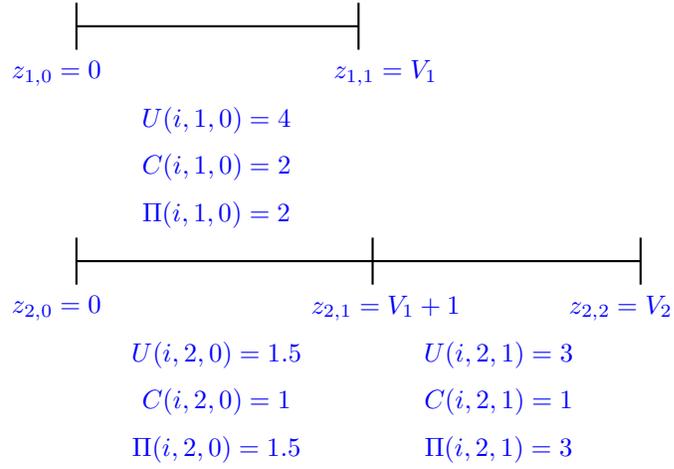
\begin{figure}[h] 
\begin{center} 
\psset{unit=1.25cm}
\begin{pspicture}(-1, -1)(4, 4)

\psline(0.25,3.5)(3.25,3.5) 
\psline(0.25,3.25)(0.25,3.75)
\psline(3.25,3.25)(3.25,3.75) 
\rput(0,3){ \small \color{blue} $z_{1,0}=0$}
\rput(3.5,3){ \small \color{blue} $z_{1,1}=V_1$}
\rput(1.7,2.5){ \small \color{blue} $U(i,1,0)=4$}
\rput(1.7,2.0){ \small \color{blue} $C(i,1,0)=2$}
\rput(1.7,1.5){ \small \color{blue} $\Pi(i,1,0)=2$}

\psline(0.25,1)(6.25,1) 
\psline(0.25,0.75)(0.25,1.25)
\psline(3.4,0.75)(3.4,1.25) 
\rput(0,0.5){ \small \color{blue} $z_{2,0}=0$}
\rput(3.5,0.5){ \small \color{blue} $z_{2,1}=V_1+1$}
\rput(1.7,0){ \small \color{blue} $U(i,2,0)=1.5$}
\rput(1.7,-0.5){ \small \color{blue} $C(i,2,0)=1$}
\rput(1.7,-1){ \small \color{blue} $\Pi(i,2,0)=1.5$}
\psline(6.25,0.75)(6.25,1.25) 
\rput(6,0.5){ \small \color{blue} $z_{2,2}=V_2$}
\rput(4.7,0){ \small \color{blue} $U(i,2,1)=3$}
\rput(4.7,-0.5){ \small \color{blue} $C(i,2,1)=1$}
\rput(4.7,-1){ \small \color{blue} $\Pi(i,2,1)=3$}

\end{pspicture}
\end{center} 
\caption{Let $B_i=2(V_1-1)$ and $V_2 \geq 2V_1-2$. 
Algorithm \ref{algogbs} outputs ($B_{i,1}=B_i, B_{i,2}=0, V_{i,1}=V_1-1, V_{i,2}=0$) with total payoff of
$4V_1-4$ to the advertiser. However, the splitting ($B_{i,1}=0, B_{i,2}=B_i, V_{i,1}=0, V_{i,2}=2V_1-2$) 
gives a total payoff of $4.5V_1-7.5 > 4V_1-4$ when $V_1>7$.} 
\label{sopt1}
\end{figure}

\begin{figure}[h] 
\begin{center} 
\psset{unit=1.25cm}
\begin{pspicture}(-1, -1)(4, 4)

\psline(0.25,3.5)(3.25,3.5) 
\psline(0.25,3.25)(0.25,3.75)
\psline(3.25,3.25)(3.25,3.75) 
\rput(0,3){ \small \color{blue} $0$}
\rput(3.5,3){ \small \color{blue} $20$}
\rput(1.7,2.5){ \small \color{blue} $U(i,1,0)=3.36$}
\rput(1.7,2.0){ \small \color{blue} $C(i,1,0)=0.8$}
\rput(1.7,1.5){ \small \color{blue} $\Pi(i,1,0)=4.2$}

\psline(0.25,1)(6.25,1) 
\psline(0.25,0.75)(0.25,1.25)
\psline(3.4,0.75)(3.4,1.25) 
\rput(0,0.5){ \small \color{blue} $0$}
\rput(3.5,0.5){ \small \color{blue} $10$}
\rput(1.7,0){ \small \color{blue} $U(i,2,0)=2$}
\rput(1.7,-0.5){ \small \color{blue} $C(i,2,0)=0.4$}
\rput(1.7,-1){ \small \color{blue} $\Pi(i,2,0)=5$}
\psline(6.25,0.75)(6.25,1.25) 
\rput(6,0.5){ \small \color{blue} $30$}
\rput(4.7,0){ \small \color{blue} $U(i,2,1)=3.2$}
\rput(4.7,-0.5){ \small \color{blue} $C(i,2,1)=0.8$}
\rput(4.7,-1){ \small \color{blue} $\Pi(i,2,1)=4$}

\end{pspicture}
\end{center} 
\caption{Greedy fractional knapsack algorithm may not return a stable solution. $B_i=12$.
This algorithm allocates all the budget to the keyword $2$ as the effective marginal payoffs is 
$\frac{2 \times 10+ 3.2 \times 10}{0.4 \times 10+ 0.8 \times 10}=\frac{52}{12}$ which is greater 
than that of keyword $1$ which is $\frac{3.36 \times 15}{0.8 \times 15} = \frac{50.4}{12}$.
However, the edge $(i,1)$ is now unstable as $MP_{i,1}^{+}= 4.2 > 4 = MP_{i,2}^{-}$. } 
\label{fksopt}
\end{figure}
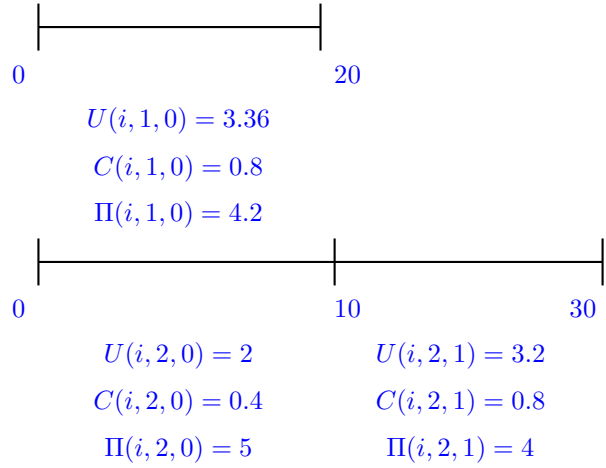

One might also like to consider another natural strongly polynomial time  greedy algorithm
based on the algorithm for {\it fractional knapsack problem}, considering each 
keyword as an item, total payoff (from all the queries that can be bought within the budget constraint)
as the value and the corresponding total cost as the size, sorting the keywords by {\it effective} 
marginal payoffs i.e. $\frac{\mbox{total payoff}}{\mbox{total cost}}$, and greedily selecting the keywords until budget is 
exhausted.
 However, it can be shown (Figure \ref{fksopt}) that this greedy approach does not always lead to 
a locally best response of an advertiser i.e. there are examples where the solution given by this algorithm 
is not stable and the advertiser can improve her payoff by local deviation.
Further, it is not clear whether some
readjustment procedure as in the {\em GREEDY READJUSTMENT PHASE} of the Algorithm \ref{algogbs}
could be applied to make such solutions stable. In this paper, we have not explored this direction in details 
and it might be interesting to make this algorithm work, if possible, by some suitable readjustment,
and compare its performance with to that of Algorithm \ref{algogbs}. For an initial insight, 
first note that, in the example given in Figure \ref{sopt1}, the marginal payoffs 
for a keyword are increasing with the partition number i.e. $\Pi(i,2,1) > \Pi(i,2,0)$.
The effective marginal payoff for keyword $2$ is $\frac{4.5V_1-7.5}{2V_1-2} >2 $ when $V_1 >7$
and this algorithms therefore spends all budget on keyword $2$, giving a payoff better than 
Algorithm \ref{algogbs}. Also, it is a stable solution because $MP_{i,2}^{-}=3 > 2 =MP_{i,1}^{+}$.
In fact, in this particular example it is the global optimum. In general, clearly when this  
algorithm returns a stable solution without a need for readjustment, it always chooses 
a better budget splitting than Algorithm \ref{algogbs},
however as we give an example below in the Figure \ref{fksopt}, it might not always return a stable solution.

\clearpage

\subsection{Approximate Nash Equilibrium($\epsilon$-NE)}
\label{ene}
Let us first define the approximate Nash equilibrium for our setting as follows.

\begin{defn}
Given a {\em BMG} $G=(\mathcal{N},\mathcal{M},\mathcal{E}$), an \emph{$\epsilon$-NE} for $G$ is defined 
as the matrix of query values $\{V_{i,j}\}$ and the budget splitting values $\{B_{i,j}\}$ such that for all $i$,
 $\sum_{(i,j) \in \mathcal{E}} \tilde{U}(i,j,V_{i,j})  \geq (1-\epsilon)  \sum_{(i,j) \in \mathcal{E}} \tilde{U}(i,j,\tilde{V}_{i,j})$
for all alternative strategy choices $\{\tilde{V}_{i,j}\}$ of advertiser $i$ that satisfies the constraints in 
Equation \ref{budcons}. 
\end{defn}

Now given the $\{V_{n,j}\}$ and $\{B_{n,j}\}$ values for all $n \in \mathcal{N}-\{i\}$, 
the approximate best response problem (\emph{ $\epsilon$-AdBRP})
for $i$ is to compute a set of $\{V_{i,j}\}$ and $\{B_{i,j}\}$ values 
satisfying the conditions in the above definition of {\em $\epsilon$-NE}. \\

As we mentioned earlier, the \emph{AdBRP} is a variant of knapsack problem and we also know that
the later can be approximated very well. Further, if we expect to get an \emph{FPTAS} 
for \emph{AdBRP}, we would expect that it also admits a pseudo-polynomial time algorithm 
(i.e. polynomial in $N$, $M$, $K$, $V_j$'s, and $B_i$'s)  as well \cite{vazapproxbook,PSbook}. 
Furthermore, all known pseudo-polynomial time algorithms for \emph{NP}-hard problems are based 
on dynamic programming \cite{vazapproxbook,PSbook}. Therefore,  naturally 
we try a similar approach.  Perhaps not surprisingly, we design a  pseudo-polynomial time algorithm for
\emph{AdBRP}. However, unlike the case of standard knapsack problems\cite{IK75,Lawler79,vazapproxbook,PSbook}, 
this algorithm does not immediately 
give a \emph{FPTAS} due to difficulty in handling the volume of queries and because for a given keyword (i.e. the item)
queries may have different costs and utilities (i.e. all units of an item are not equivalent in cost and value) etc.
Nevertheless, by judiciously utilizing the properties of the optimum and a double-layered approximation we will 
indeed present a \emph{FPTAS}  for  \emph{AdBRP} (Theorem \ref{enealgo}).
We will devote the Sections \ref{appseudopolyalgo}, \ref{appseudopolyapprox}, \ref{apfptas} on developing a FPTAS for AdBRP.

\begin{theo}
\label{enealgo}
There is a FPTAS for AdBRP.
\end{theo}

Thus, \textit{$\epsilon$-NE} is indeed another reasonable solution concept for the Broad-Match Game.

\subsubsection{A Pseudo Polynomial Time Algorithm for AdBRP}
\label{appseudopolyalgo}

Without loss of generality, first let us assume that   
all relevant parameters such as budget, utilities, and costs are integer valued. 
This can be achieved by suitable approximation to rational numbers and then by appropriate 
multiplication factors, without significant increase in instance size.  \\

Let $P$ be the maximum total utility that the advertiser $i$ can derive from a {\it single} keyword i.e. 
from any one of the $j \in \mathcal{M}$ while respecting her budget constraint. 
Therefore, $MP$ is a trivial upperbound on the total utility that can be achieved by any solution (i.e. any choice of 
feasible $x_{i,l}$'s). 
For $j \in \mathcal{M}$ and $p=1,2,\dots, MP$ let us define, 
\begin{equation}
A(j,p)= \left\{ \begin{array}{l}
 \textrm{ {\em Min}} ~~ \sum_{l=1}^j \tilde{C}(i,l,x_{i,l})  \\
\\
   \textrm{ {\em s.t.}   }  ~~ \sum_{l=1}^j\tilde{U}(i,l,x_{i,l}) \geq p  \\
     ~~~~~~~~0 \leq  x_{i,l} \leq V_l  ~~~\textrm{  for  } l=1,2,\dots, j  \\
 ~~~~~~~~ x_{i,l} \in \mathcal{Z} ~~~ \textrm{  for  } l=1,2,\dots, j \\  
 \\
\infty  ~~~ \textrm{ if no solution to the above minimization problem exists} 
\end{array} \right. 
\end{equation}

Clearly, the values $A(1,p)$ for all $p=1,2,\dots, MP$ can each be efficiently computed ( in 
$O(N log(V_1))$ time) by utilizing the partition structure as per Lemma \ref{par1} and
the Equations \ref{utfun} and \ref{ctfun}. The term $log(V_1)$ instead of $V_1$ comes 
from the the fact that it is enough to perform a binary search inside a partition for an appropriate $x_{i,1}$. 
Therefore, the values $A(1,p)$ for all $p=1,2,\dots, MP$ can be together computed
in $O(N log(V_1) MP )$ time. Let $V=\sum_{j \in \mathcal{M}} V_j$.  
Now, $A(j,p)$ for all $j=2,3,\dots,M$ and $p=1,2,\dots, MP$ can be computed together in $O(M^2P N V )$ using 
following recurrence relation. Further, for each choice of $j=1, 2,3,\dots,M$ and $p=1,2,\dots, MP$, we 
also compute and store the set of $x_{i,j}$'s values that achieves the value of $A(j,p)$, and this 
task can be performed in same order of time complexity.   

\begin{eqnarray}
A(j+1,p) &=& \left\{ \begin{array}{l}
 \textrm{ {\em Min}} ~~ \sum_{l=1}^j \tilde{C}(i,l,x_{i,l})  + \tilde{C}(i,j+1,x_{i,j+1})   \\
\\
   \textrm{ {\em s.t.}   }  ~~ \sum_{l=1}^j\tilde{U}(i,l,x_{i,l}) + \tilde{U}(i,j+1,x_{i,j+1}) \geq p  \\
     ~~~~~~~~0 \leq  x_{i,l} \leq V_l  ~~~\textrm{  for  } l=1,2,\dots, j+1  \\
 ~~~~~~~~ x_{i,l} \in \mathcal{Z} ~~~ \textrm{  for  } l=1,2,\dots, j+1\\  
 \\
\infty  ~~~ \textrm{ if no solution to the above minimization problem exists} 
\end{array} \right. \nonumber 
 \\
\nonumber \\
\nonumber \\
  & = & \min_{x_{i,j+1} \in \{0,1,2,\dots,V_{j+1}\}: \tilde{U}(i,j+1,x_{i,j+1}) \leq p} 
\left\{ \tilde{C}(i,j+1,x_{i,j+1})+ A(j, p-\tilde{U}(i,j+1,x_{i,j+1})) \right\} \label{dtabcal}
\end{eqnarray}

After computing the $A(j,p)$'s,  we can obtain the solution of AdBRP and the optimum value via 
\begin{equation}
OPT = \max_{p=1,2,\dots,MP} \left\{ p : A(M,p) \leq B_i\right\}. 
\end{equation}

Thus, we have a pseudo-polynomial time algorithm for
\emph{AdBRP}.

\subsubsection{A Pseudo Polynomial Time Approximation Scheme for AdBRP}
\label{appseudopolyapprox}

Now, we will convert the pseudo polytime algorithm of Section \ref{appseudopolyalgo} to an approximation scheme  
by appropriately rounding the payoffs of the advertisers. Let us call this scheme 
\textbf{AS1}. \\

For a given $\epsilon > 0$, let $T = \frac{\epsilon P}{M}$ and 
let us round the utility functions from  $\tilde{U}(i,l,x_{i,l})$ to 
$\tilde{U}^{'}(i,l,x_{i,l})  = \lfloor\frac{\tilde{U}(i,l,x_{i,l})}{T}\rfloor$. 
Now, let us use the above dynamic programming algorithm by replacing the upper bound on utility 
$P$ to $\lfloor\frac{P}{T}\rfloor$ and utilities  $\tilde{U}(i,l,x_{i,l})$
to $\tilde{U}^{'}(i,l,x_{i,l})$. Let us denote the solution returned as $V_{i,j}^{'}$ i.e. the 
optimum of the problem after above rounding is attained at $x_{i,j} = V_{i,j}^{'}, j \in \mathcal{M}$. 
Correspondingly, let the optimal solution without any rounding is $V_{i,j}, j \in \mathcal{M}$. 
Note that  
the values $\{V_{i,j}\}$ are still feasible for the rounded problem. \\

Now since $\{V_{i,j}^{'}\}$ is optimal for the rounded problem 
we have 
\begin{eqnarray}
\sum_{j \in \mathcal{M}} \tilde{U}^{'}(i,j,V_{i,j}^{'}) &\geq&  \sum_{j \in \mathcal{M}} \tilde{U}^{'}(i,j,V_{i,j}) \nonumber\\
& \geq & \sum_{j \in \mathcal{M}} \left( \frac{\tilde{U}(i,j,V_{i,j})}{T} - 1 \right) \nonumber \\
& = & \sum_{j \in \mathcal{M}} \frac{\tilde{U}(i,j,V_{i,j})}{T} - M  \label{ineq1} 
\end{eqnarray}

Further, using the inequality \ref{ineq1}, we can get 
\begin{eqnarray*}
\sum_{j \in \mathcal{M}} \tilde{U}(i,j,V_{i,j}^{'}) & \geq & \sum_{j \in \mathcal{M}}T \tilde{U}^{'}(i,j,V_{i,j}^{'})\\
& \geq & \sum_{j \in \mathcal{M}} \tilde{U}(i,j,V_{i,j}) - TM  \\
& = &  \sum_{j \in \mathcal{M}} \tilde{U}(i,j,V_{i,j}) - \epsilon P \\
& \geq & (1 - \epsilon) \sum_{j \in \mathcal{M}} \tilde{U}(i,j,V_{i,j})
\end{eqnarray*}
where the last inequality is implied by the fact that the optimal value without rounding will 
at least be the maximum derived from a single keyword i.e. $\sum_{j \in \mathcal{M}} \tilde{U}(i,j,V_{i,j}) \geq P$. 
Therefore, the  $\{ V_{i,j}^{'} \}$ gives an $(1-\epsilon)$ approximation
to the advertiser $i$'s best response. The time complexity of the dynamic programming applied to obtain 
the solution  $\{ V_{i,j}^{'} \}$ is 
$O(M^2\lfloor\frac{P}{T}\rfloor N V ) = O(M^2N (\lfloor\frac{M}{\epsilon}\rfloor)V)$ which is polynomial 
in $N$, $M$, $V$ and $\frac{1}{\epsilon}$.
Note that this scheme is still pseudo polynomial 
due to appearance of $V$.
This terms appears because when we build up the dynamic programming table we need to use 
the Equation \ref{dtabcal}.  
Nevertheless, this algorithm will be helpful in designing our FPTAS.   \\

\subsubsection{FPTAS for AdBRP}
\label{apfptas}
Now using the approximation scheme {\bf AS1} and by judiciously truncating set of possible query values 
we will present an FPTAS for AdBRP.  
Let us first note that for each keyword only the number of queries, 
starting the first one, that can be bought under budget constraint $B_i$ are 
feasible. Therefore, for the ease of notation, without loss of generality, 
we can assume that $V_j$ is infact the maximum number of queries 
that can be bought under budget constraint meaning all possible queries of a particular keyword are feasible if $i$ wanted to spend her total
budget on this keyword. Let $V=\sum_{j \in M}V_j$. Also, without loss of generality we have $U(i,j,\lambda) > 0$ for all $\lambda$ and $j$ such that
$(i,j) \in \mathcal{E}$. \\ 

Now, we perform a first layer of approximation by judiciously truncating the set of
possible query values to obtain Lemma \ref{fptasfinally}.  Lemma \ref{optprop1} is just a warmup. 

\begin{lm}
\label{optprop1}
There is a feasible solution $\{\hat{V}_{i,j}\}$ of \textbf{AdBRP} with the total utility $\hat{OPT}$ satisfying the following properties
where $OPT$ is the total utility for the optimal solution of \textbf{AdBRP}:

\begin{itemize}
\item At most for \textbf{one} $j \in M$, $\hat{V}_{i,j} \notin \{z_{j,0},z_{j,1}, z_{j,2}, \dots, z_{j,\Lambda_j}\}$

\item   $\hat{OPT} \geq (1-\delta) OPT$ where $\delta = \max_{j,\lambda} \frac{U(i,j,\lambda)}{OPT}$.

\end{itemize} 
\end{lm}
\proof 

Consider all the keywords $j$ for which  $V_{i,j} \notin \{z_{j,0},z_{j,1}, z_{j,2}, \dots, z_{j,\Lambda_j}\}$.
Now among these keywords, for the keywords with the maximum value of $\pi(i,j,V_{i,j})$, move the budget from one to the other 
unless all keywords of such type except one satisfies the above condition, this moving around of budgets among the keywords with the 
same value of $\pi(i,j,V_{i,j})$ does not change the total utility. Thus, we can safely assume that there is a single keyword
satisfying  $V_{i,j} \notin \{z_{j,0},z_{j,1}, z_{j,2}, \dots, z_{j,\Lambda_j}\}$ with the maximum value of $\pi(i,j,V_{i,j})$ and say its $l$. 
Further, we can move budget from all other keywords satisfying $V_{i,j} \notin \{z_{j,0},z_{j,1}, z_{j,2}, \dots, z_{j,\Lambda_j}\}$
to $l$ until they satisfy $V_{i,j} \in \{z_{j,0},z_{j,1}, z_{j,2}, \dots, z_{j,\Lambda_j}\}$,
while losing at most $u(i,l,V_{i,l})$. This is because while doing this reallocation we lose only when we 
do not have enough budget to move to buy the next query of $l$. \qed.

\begin{lm}
\label{fptasfinally}
For every given parameter $\epsilon < 1$, there is a feasible solution $\{\hat{V}_{i,j}\}$ of 
\textbf{AdBRP} with the total utility $\hat{OPT}$ satisfying the following properties
where $OPT$ is the total utility for the optimal solution of \textbf{AdBRP}:

\begin{itemize}

\item   $\hat{OPT} \geq (1-\epsilon^2) OPT$ 

\item $\hat{V}_{i,j} \in  \{ y_{j,0}, y_{j,1}, y_{j,2}, \dots ,  y_{j,n_j}\}$ for all $j \in M$
where the $y_{j,.}$ values are defined 
as follows: \\

for each partition $\lambda =1,2,\dots,\Lambda_j$ of the keyword $j$  
let $z_{j,\lambda}-z_{j,\lambda-1} = a \lceil \frac{M}{\epsilon^2} \rceil + b$ where $a$ and $b$ are non-negative integers
and $b < \lceil \frac{M}{\epsilon^2}\rceil$ (recall that $z_{j,\lambda}-z_{j,\lambda-1}$ is the size of partition $\lambda$). Now  
divide the partition $\lambda$ in to $min \{a,1\} . \lceil \frac{M}{\epsilon^2}\rceil + b$ sub-partitions, the first 
$min \{a,1\} . \lceil \frac{M}{\epsilon^2}\rceil$ sub-partitions being of size
 $a$ queries each and the next $b$ partitions 
each of size $1$ query each. Define $y_{j,0}=0$ and $ y_{j,1}, y_{j,2}, \dots, y_{j,n_j}$ as the end points of the 
sub-partitions created as above. 
\end{itemize} 
\end{lm}
\proof 
Clearly, each partition is divided in to at most $2 \lceil \frac{M}{\epsilon^2}\rceil$ 
sub-partitions and since $\Lambda_j = O(N)$ we obtain $n_j = O(\frac{NM}{\epsilon^2})$. 

Let $\{V_{i,j}\}$ be an optimal solution for \textbf{AdBRP}, then we find a solution  $\{\hat{V}_{i,j}\}$ as 
follows: 

\begin{itemize}
\item if $V_{i,j} \in  \{ y_{j,0}, y_{j,1}, y_{j,2}, \dots ,  y_{j,n_j}\}$ then $\hat{V}_{i,j}=V_{i,j}$, therefore 
we do not lose anything in total utility coming from keyword $j$ by moving from $V_{i,j}$ to $\hat{V}_{i,j}$. 

\item if $V_{i,j} \notin  \{ y_{j,0}, y_{j,1}, y_{j,2}, \dots ,  y_{j,n_j}\}$ then define $\hat{V}_{i,j}$ to be
the maximum value among $\{ y_{j,0}, y_{j,1}, y_{j,2}, \dots ,  y_{j,n_j}\}$ which is smaller than $V_{i,j}$. 
Note that this case arises only when $V_{i,j}$ comes from a sub-partition of size $a > 1$ and in that case
there are at least $\lceil \frac{M}{\epsilon^2}\rceil$ sub-partitions of
 size $a$ with the same utility value $u(i,j,V_{i,j})$
for each query in those sub-partitions. The maximum utility that we can lose 
by truncating $V_{i,j}$ to $\hat{V}_{i,j}$
is 
$\leq a ~ u(i,j,V_{i,j}) \leq \frac{1}{\lceil \frac{M}{\epsilon^2}\rceil} 
\left( \lceil \frac{M}{\epsilon^2}\rceil ~a ~ u(i,j,V_{i,j}) \right) \leq \frac{1}{\lceil \frac{M}{\epsilon^2}\rceil} OPT
\leq \frac{\epsilon^2}{M} OPT$.

Note that the way we have constructed $\{\hat{V}_{i,j}\}$, it satisfies the budget constraint of the advertiser $i$
because $\hat{V}_{i,j} \leq V_{i,j}$ and hence $\{\hat{V}_{i,j}\}$ is a feasible solution of \textbf{AdBRP}. 
Now, across all the keywords we can lose at most $M ~ \frac{\epsilon^2}{M} OPT$ i.e. $\epsilon^2 OPT$ by truncating 
the optimal solution $\{V_{i,j}\}$ to the feasible solution $\{\hat{V}_{i,j}\}$. \qed. 
\end{itemize}

Now we perform a second layer of approximation by applying \textbf{AS1} with truncated set of possible
query values as per Lemma \ref{fptasfinally}.
We present this approximation algorithm called \textbf{AS2} in the following.

\begin{enumerate}

\item Divide each partition of each keyword in to several sub-partitions as follows 
(same as in the proof of Lemma \ref{fptasfinally}). 
 For each partition $\lambda =1,2,\dots,\Lambda_j$ of the keyword $j$  
let $z_{j,\lambda}-z_{j,\lambda-1} = a \lceil \frac{M}{\epsilon^2} \rceil + b$ where $a$ and $b$ are non-negative integers
and $b < \lceil \frac{M}{\epsilon^2}\rceil$ (recall that $z_{j,\lambda}-z_{j,\lambda-1}$ is the size of partition $\lambda$). Now  
divide the partition $\lambda$ in to $min \{a,1\} . \lceil \frac{M}{\epsilon^2}\rceil + b$ sub-partitions, the first 
$min \{a,1\} . \lceil \frac{M}{\epsilon^2}\rceil$ sub-partitions being of size
 $a$ queries each and the next $b$ partitions 
each of size $1$ query each. Define $y_{j,0}=0$ and $ y_{j,1}, y_{j,2}, \dots, y_{j,n_j}$ as the end points of the 
sub-partitions created as above. 

\item Apply \emph{AS1} by restricting the query values to take 
 values only from the set $\{ y_{j,0}, y_{j,1}, y_{j,2}, \dots, y_{j,n_j} \}$
i.e. by changing the recurrence relation \ref{dtabcal} to 
\begin{equation}
A(j+1,p) = \min_{x_{i,j+1} \in \{y_{j,0}, y_{j,1}, y_{j,2}, \dots,  y_{j,n_j}\}: \tilde{U}(i,j+1,x_{i,j+1}) \leq p} 
\left\{ \tilde{C}(i,j+1,x_{i,j+1})+ A(j, p-\tilde{U}(i,j+1,x_{i,j+1})) \right\} \label{dtabcaltruncate}
\end{equation}
and with the error parameter $\tilde{\epsilon}$, where  
$\frac{1}{\tilde{\epsilon}} = \frac{1}{\epsilon}+1 \leq \frac{2}{\epsilon}$.

\end{enumerate}

The total utility of the solution returned by the algorithm \textbf{AS2} is 
\begin{equation*}
\geq (1-\tilde{\epsilon})(1-\epsilon^2) OPT = (1-\frac{\epsilon}{1+\epsilon})(1-\epsilon^2) OPT = (1-\epsilon) OPT.
\end{equation*} 

The time complexity in constructing the sub-partitions is $O(\frac{NM}{\epsilon^2})$ and that of applying 
\emph{AS1} on the truncated set of query values $\{ y_{j,0}, y_{j,1}, y_{j,2}, \dots, y_{j,n_j} \}$
is $ O(M^2N (\lfloor\frac{M}{\tilde{\epsilon}}\rfloor) \frac{NM}{\epsilon^2} ) = O(\frac{M^4N^2}{\epsilon^3})$ as 
$\frac{1}{\tilde{\epsilon}} = \frac{1}{\epsilon}+1 \leq \frac{2}{\epsilon} = O(\frac{1}{\epsilon}) $. 
Therefore, \textbf{AS2} is indeed a \textbf{FPTAS} for \textbf{AdBRP}.

\subsection{To Broad-match or not to Broad-match}
\label{bmnbm}

In this section, we start out with a very important observation in the following theorem.

\begin{theo}
\label{acdilm}
A {\em BMG} does not necessarily have a unique {\em BME} and the different {\em BME}s can yield different 
revenues to the auctioneer. Moreover, in one of them the auctioneer loses while in the other one it gains in terms of revenue.
\end{theo}

The Theorem \ref{acdilm} leaves the auctioneer in a dilemma about whether he should 
broad-match or not. If he could somehow predict which choice of broad match
lead to a revenue improvement for him and which not, he could potentially 
choose the ones leading to a revenue improvement. 
This brings 
forth one of the big questions left open in this paper, 
that is of efficiently computing a  \textit{BME} / \textit{$\epsilon$-NE}, if one exists, 
given a choice of \textit{broad match}. We plan to explore it in our future works.
The proof of the above theorem follows from examples constructed in Figures \ref{twobme0}, \ref{twobme1}, \ref{twobme2}.  
Further, {\bf in all the examples we take $K=2, \gamma_1=1, \gamma_2=0.7$.}  \\

\begin{figure}[h] 
\begin{center} 
\psset{unit=1.25cm}
\begin{pspicture}(0, -3)(4, 6)

\psline(0.25,-3)(3.75,-2) 
\rput{20}(2,-2.3){ \small \color{red} 2}

\pscircle[linecolor=blue](0,-3){0.25}
\rput(0,-3){\small \color{blue} 6}
\rput(-0.4,-3.4){\small \color{red} $B_6 > 0 $}

\pscircle[linecolor=green](4,-2){0.25} 
\rput(4,-2){\small \color{green} 3}
\rput(4,-2.4){\small \color{red} $V_3$}

\psline(0.25,-1)(3.75,-2) 
\rput{-20}(2,-1.3){\small \color{red} 3}

\pscircle[linecolor=blue](0,-1){0.25} 
\rput(0,-1){\small \color{blue} 5}
\rput(-0.4,-1.4){\small \color{red} $B_5=0.6V_3$}

\psline(0.25,0)(3.75,1) 
\rput{20}(1,0.4){ \small \color{red} 1.5}

\pscircle[linecolor=blue](0,0){0.25} 
\rput(0,0){\small \color{blue} 4}
\rput(-0.4,0.4){\small \color{red} $B_4>0 $}

\pscircle[linecolor=green](4,1){0.25} 
\rput(4,1){\small \color{green} 2}
\rput(4,0.6){\small \color{red} $V_2$}

\psline(0.25,2)(3.75,1) 
\rput{-20}(2.5,1.6){\small \color{red} 5}

\pscircle[linecolor=blue](0,2){0.25} 
\rput(0,2){\small \color{blue} 3}
\rput(-0.4,1.6){\small \color{red} $B_3=2.25V_2$}

\psline(0.25,3)(3.75,4) 
\rput{20}(2,3.7){\small \color{red} 3}

\pscircle[linecolor=blue](0,3){0.25} 
\rput(0,3){\small \color{blue} 2}
\rput(-0.4,3.4){ \small \color{red} $B_2=0.6 V_3$}

\pscircle[linecolor=green](4,4){0.25} 
\rput(4,4){\small \color{green} 1}
\rput(4,3.6){\small \color{red} $V_1$}

\psline(0.25,5)(3.75,4) 
\rput{-20}(2,4.7){ \small \color{red} 5}

\pscircle[linecolor=blue](0,5){0.25} 
\rput(0,5){\small \color{blue} 1}
\rput(-0.6,4.6){\small \color{red} $B_1 = 2.3V_1$}

\rput(4, - 3){\color{orange} $R_0=0.9V_1+0.45V_2+0.6V_3$} 
\rput(-0.4, 6){ \color{cyan} Advertisers}
\rput(4, 6){ \color{cyan} Keywords}

\rput(4, -3.5){\color{blue} $E_0=7.1V_1+6.05V_2+4.4V_3$ }

\psline[linestyle=dashed](0.25,3)(3.75,-2) 
\rput{-45}(3.5, - 1){\small \color{red} 3.4}

\psline[linestyle=dashed](0.25,0)(3.75,4) 
\rput{45}(2.5, 2.9){\small \color{red} 2}

\psline[linestyle=dashed](0.25,-1)(3.75,1) 
\rput{45}(1.2, -0.2){\small \color{red} 4}

\end{pspicture}
\end{center} 
\caption{A {\em BMG} (without dashed edges) and its extension (with dashed edges). The values shown along the edges
are $s_{i,j}$ respectively. Further, note that it is a good broad-match 
as valuations $s_{i,j}$ of an advertiser along new edges is greater than her valuation along old edges. } 
\label{twobme0}
\end{figure}
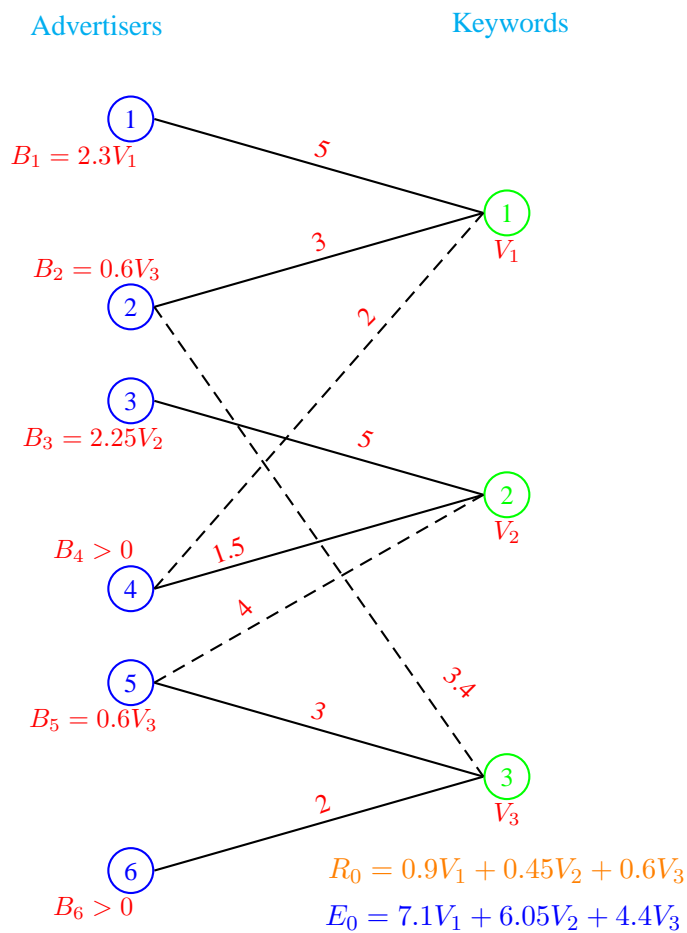

\begin{figure}[h] 
\begin{center} 
\psset{unit=1.25cm}
\begin{pspicture}(0, -3)(4, 6)

\psline(0.25,-3)(3.75,-2) 
\rput{20}(2,-2.3){ \small \color{red} $V_3$}

\pscircle[linecolor=blue](0,-3){0.25} 
\rput(0,-3){\small \color{blue} 6}

\pscircle[linecolor=green](4,-2){0.25} 
\rput(4,-2){\small \color{green} 3}

\psline(0.25,-1)(3.75,-2) 
\rput{-20}(2,-1.3){\small \color{red} 0}

\pscircle[linecolor=blue](0,-1){0.25} 
\rput(0,-1){\small \color{blue} 5}

\psline(0.25,0)(3.75,1) 
\rput{20}(1,0.4){ \small \color{red} $V_2$}

\pscircle[linecolor=blue](0,0){0.25} 
\rput(0,0){\small \color{blue} 4}

\pscircle[linecolor=green](4,1){0.25} 
\rput(4,1){\small \color{green} 2}

\psline(0.25,2)(3.75,1) 
\rput{-20}(2.5,1.6){\small \color{red} $V_2$}

\pscircle[linecolor=blue](0,2){0.25} 
\rput(0,2){\small \color{blue} 3}

\psline(0.25,3)(3.75,4) 
\rput{20}(2,3.7){\small \color{red} 0}

\pscircle[linecolor=blue](0,3){0.25} 
\rput(0,3){\small \color{blue} 2}

\pscircle[linecolor=green](4,4){0.25} 
\rput(4,4){\small \color{green} 1}

\psline(0.25,5)(3.75,4) 
\rput{-20}(2,4.7){ \small \color{red} $V_1$}

\pscircle[linecolor=blue](0,5){0.25} 
\rput(0,5){\small \color{blue} 1}

\rput(4, - 3){\color{orange} $R=R_0+ 11.4 \frac{V_3}{7}-0.3V_1$} 

\rput(4, -3.5){\color{blue} $E=E_0+1.4V_3 - 0.7V_1$ }

\psline[linestyle=dashed](0.25,3)(3.75,-2) 
\rput{-45}(3.5, - 1){\small \color{red} $V_3$}

\psline[linestyle=dashed](0.25,0)(3.75,4) 
\rput{45}(2.5, 2.9){\small \color{red} $V_1$}

\psline[linestyle=dashed](0.25,-1)(3.75,1) 
\rput{45}(1.2, -0.2){\small \color{red} $\frac{4}{7}V_3$}

\rput(-1.5,3.25){\small  $MP_{2,1}^{+}=0.5$}
\rput(-1.5,2.65){\small  $MP_{2,3}^{-}= 4.67$}
\rput(-1.5,-0.75){\small  $MP_{5,2}^{-}=1.67$}
\rput(-1.5,-1.35){\small  $MP_{5,3}^{+}= 0.5$}

\end{pspicture}
\end{center} 
\caption{One {\em BME} for the extension of {\em BMG} in Figure \ref{twobme0} (
the values  $V_{i,j}$ at this {\em BME} are shown along corresponding edges). \newline
\textbf{Not a pure Nash equilibrium but an $\epsilon$-NE for $\epsilon \geq 0.15$}. } 
\label{twobme1}
\end{figure}
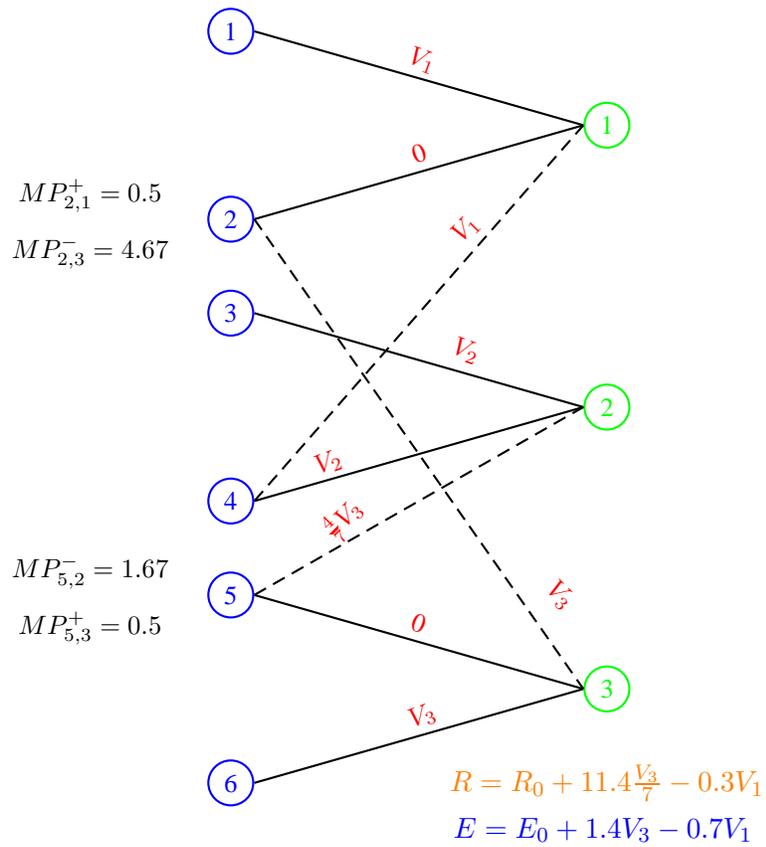

\begin{figure}[h] 
\begin{center} 
\psset{unit=1.25cm}
\begin{pspicture}(0, -3)(4, 6)

\psline(0.25,-3)(3.75,-2) 
\rput{20}(2,-2.3){ \small \color{red} $V_3$}

\pscircle[linecolor=blue](0,-3){0.25} 
\rput(0,-3){\small \color{blue} 6}

\pscircle[linecolor=green](4,-2){0.25} 
\rput(4,-2){\small \color{green} 3}

\psline(0.25,-1)(3.75,-2) 
\rput{-20}(2,-1.3){\small \color{red} $V_3$ }

\pscircle[linecolor=blue](0,-1){0.25} 
\rput(0,-1){\small \color{blue} 5}

\psline(0.25,0)(3.75,1) 
\rput{20}(1,0.4){ \small \color{red} $V_2$}

\pscircle[linecolor=blue](0,0){0.25} 
\rput(0,0){\small \color{blue} 4}

\pscircle[linecolor=green](4,1){0.25} 
\rput(4,1){\small \color{green} 2}

\psline(0.25,2)(3.75,1) 
\rput{-20}(2.5,1.6){\small \color{red} $V_2$}

\pscircle[linecolor=blue](0,2){0.25} 
\rput(0,2){\small \color{blue} 3}

\psline(0.25,3)(3.75,4) 
\rput{20}(2,3.7){\small \color{red} $\frac{3}{7}V_3$}

\pscircle[linecolor=blue](0,3){0.25} 
\rput(0,3){\small \color{blue} 2}

\pscircle[linecolor=green](4,4){0.25} 
\rput(4,4){\small \color{green} 1}

\psline(0.25,5)(3.75,4) 
\rput{-20}(2,4.7){ \small \color{red} $V_1$}

\pscircle[linecolor=blue](0,5){0.25} 
\rput(0,5){\small \color{blue} 1}

\rput(4, - 3){\color{orange} $R=R_0+ 9.3 \frac{V_3}{7}-0.3V_1$} 

\rput(4, -3.5){\color{blue} $E=E_0+0.7 \frac{3V_3}{7} - 0.7V_1$ }

\psline[linestyle=dashed](0.25,3)(3.75,-2) 
\rput{-45}(3.5, - 1){\small \color{red} $0$}

\psline[linestyle=dashed](0.25,0)(3.75,4) 
\rput{45}(2.5, 2.9){\small \color{red} $V_1$}

\psline[linestyle=dashed](0.25,-1)(3.75,1) 
\rput{45}(1.2, -0.2){\small \color{red} $0$}

\rput(-1.5,3.25){\small  $MP_{2,1}^{-}=0.5$}
\rput(-1.5,2.65){\small  $MP_{2,3}^{+}= 0.48$}
\rput(-1.5,-0.75){\small  $MP_{5,2}^{+}=1.67$}
\rput(-1.5,-1.35){\small  $MP_{5,3}^{-}= 4$}

\end{pspicture}
\end{center} 
\caption{Another {\em BME} for the extension of {\em BMG} in Figure \ref{twobme0}.\newline
\textbf{Also a pure Nash equilibrium, therefore an $\epsilon$-NE for $\epsilon \geq 0$.} } 
\label{twobme2}
\end{figure}
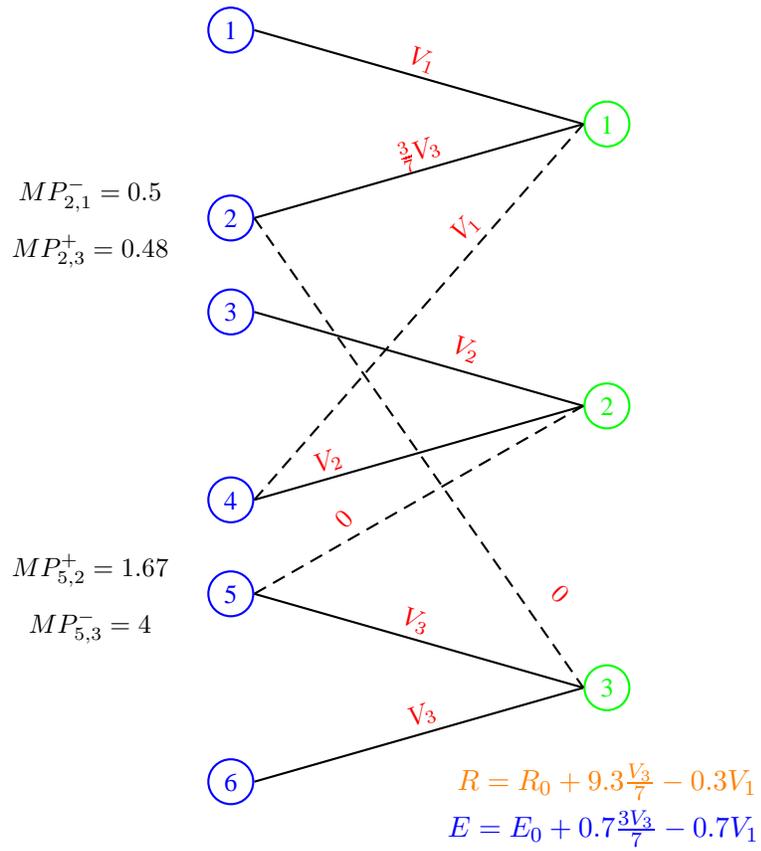

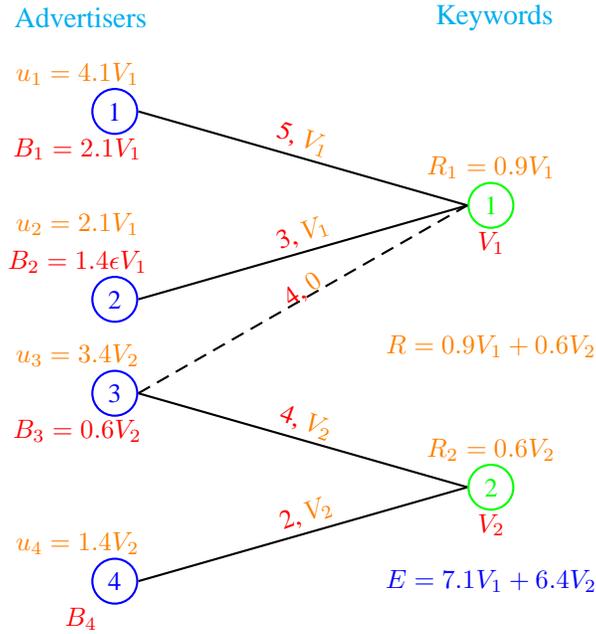
\begin{figure}[h] 
\begin{center} 
\psset{unit=1.25cm}
\begin{pspicture}(0, 0)(4, 6.5)
\psline(0.25,0)(3.75,1) 
\rput{20}(2,0.7){ \small \color{red} 2, \color{orange} $V_2$}

\pscircle[linecolor=blue](0,0){0.25}
\rput(0,0){\small \color{blue} 4}
\rput(-0.4,-0.4){ \small \color{red} $B_4$}
\rput(-0.4, 0.4) {\small \color{orange} $u_4=1.4V_2$}

\pscircle[linecolor=green](4,1){0.25} 
\rput(4,1){\small \color{green} 2}
\rput(4,0.6){\small \color{red} $V_2$}
\rput(4,1.4){\small \color{orange} $R_2=0.6V_2$}

\psline(0.25,2)(3.75,1) 
\rput{-20}(2,1.7){ \small \color{red} 4, \color{orange} $V_2$}

\pscircle[linecolor=blue](0,2){0.25} 
\rput(0,2){\small \color{blue} 3}
\rput(-0.4,1.6){\small \color{red} $B_3=0.6 V_2$}
\rput(-0.4, 2.4) {\small \color{orange} $u_3=3.4V_2$}

\psline(0.25,3)(3.75,4) 
\rput{20}(2,3.7){\small \color{red} 3, \color{orange} $V_1$}

\pscircle[linecolor=blue](0,3){0.25} 
\rput(0,3){\small \color{blue} 2}
\rput(-0.4,3.4){\small \color{red} $B_2=1.4 \epsilon V_1$}
\rput(-0.4, 3.8) {\small \color{orange} $u_2=2.1V_1$}

\pscircle[linecolor=green](4,4){0.25} 
\rput(4,4){\small \color{green} 1}
\rput(4,3.6){\small \color{red} $V_1$}
\rput(4,4.4){\small \color{orange} $R_1=0.9V_1$}

\psline(0.25,5)(3.75,4) 
\rput{-20}(2,4.7){\small \color{red} 5, \color{orange} $V_1$}

\pscircle[linecolor=blue](0,5){0.25} 
\rput(0,5){\small \color{blue} 1}
\rput(-0.4,4.6){\small \color{red} $B_1=2.1V_1$}
\rput(-0.4, 5.4) {\small \color{orange} $u_1=4.1V_1$}

\rput(4, 2.5){\small \color{orange} $R= 0.9V_1 + 0.6V_2$}
\rput(-0.4, 6){ \color{cyan} Advertisers}
\rput(4, 6){ \color{cyan} Keywords}

\psline[linestyle=dashed](0.25,2)(3.75,4) 
\rput{35}(2,3.1){\small \color{red} 4, \color{orange} $0$}

\rput(4,0){\small \color{blue} $E=7.1V_1 + 6.4V_2$}
\end{pspicture}
\end{center} 
\caption{No Shift in Equilibrium due to the new edge (3,1). The values shown along edge $(i,j)$ is $(s_{i,j},V_{i,j})$.
$u_i$ denote total payoff of advertiser $i$.} 
\label{noshift}
\end{figure}

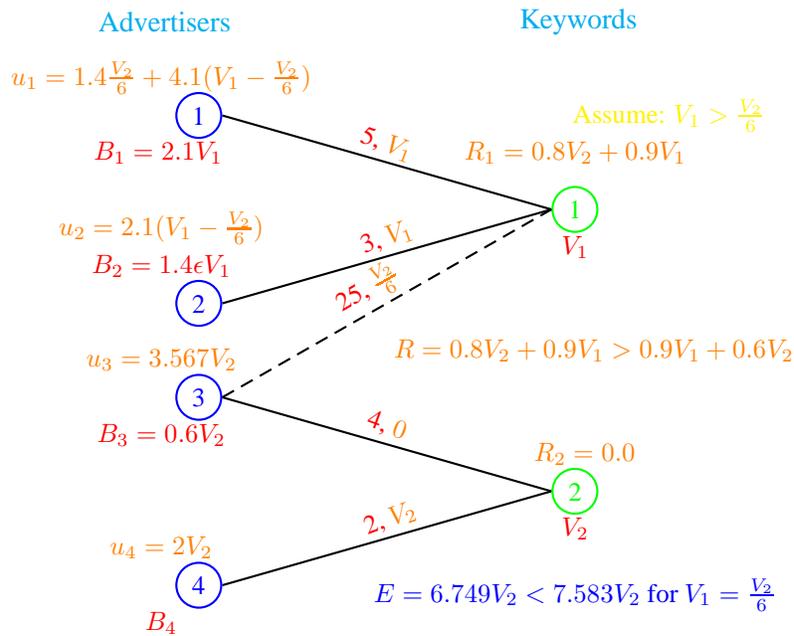
\begin{figure}[h] 
\begin{center} 
\psset{unit=1.25cm}
\begin{pspicture}(0, 0)(4, 6)
\psline(0.25,0)(3.75,1) 
\rput{20}(2,0.7){ \small \color{red} 2, \color{orange} $V_2$}

\pscircle[linecolor=blue](0,0){0.25}
\rput(0,0){\small \color{blue} 4}
\rput(-0.4,-0.4){\small \color{red} $B_4$}
\rput(-0.4, 0.4) {\small \color{orange} $u_4=2V_2$}

\pscircle[linecolor=green](4,1){0.25} 
\rput(4,1){\small \color{green} 2}
\rput(4,0.6){\small \color{red} $V_2$}
\rput(4.1,1.4){\small \color{orange} $R_2=0.0$}

\psline(0.25,2)(3.75,1) 
\rput{-20}(2,1.7){\small \color{red} 4, \color{orange} $0$}

\pscircle[linecolor=blue](0,2){0.25} 
\rput(0,2){\small \color{blue} 3}
\rput(-0.4,1.6){\small \color{red} $B_3=0.6 V_2$}
\rput(-0.4, 2.4) {\small \color{orange} $u_3=3.567V_2$}

\psline(0.25,3)(3.75,4) 
\rput{20}(2,3.7){\small \color{red} 3, \color{orange} $V_1$}

\pscircle[linecolor=blue](0,3){0.25} 
\rput(0,3){\small \color{blue} 2}
\rput(-0.4,3.4){\small \color{red} $B_2=1.4 \epsilon V_1$}
\rput(-0.4, 3.8) {\small \color{orange} $u_2=2.1(V_1-\frac{V_2}{6})$}

\pscircle[linecolor=green](4,4){0.25} 
\rput(4,4){\small \color{green} 1}
\rput(4,3.6){\small \color{red} $V_1$}
\rput(4,4.6){\small \color{orange} $R_1=0.8V_2+0.9V_1$}

\psline(0.25,5)(3.75,4) 
\rput{-20}(2,4.7){\small \color{red} 5, \color{orange} $V_1$}

\pscircle[linecolor=blue](0,5){0.25} 
\rput(0,5){\small \color{blue} 1}
\rput(-0.4,4.6){\small \color{red} $B_1=2.1V_1$ }
\rput(-0.4, 5.4) {\small  \color{orange} $u_1=1.4\frac{V_2}{6}+4.1(V_1-\frac{V_2}{6})$}

\rput(4.2, 2.5){\small \color{orange} $R=0.8V_2+0.9V_1 > 0.9V_1 + 0.6V_2$} 
\rput(-0.4, 6){ \color{cyan} Advertisers}
\rput(4, 6){ \color{cyan} Keywords}

\psline[linestyle=dashed](0.25,2)(3.75,4) 
\rput{30}(1.8,3.16){\small \color{red} 25, \color{orange} $\frac{V_2}{6}$}

\rput(5,5) {\small \color{yellow} Assume: $V_1 > \frac{V_2}{6}$}

\rput(4,- 0.1){\small \color{blue} $E= 6.749V_2 < 7.583 V_2$ for $V_1= \frac{V_2}{6}$}

\end{pspicture}
\end{center} 
\caption{A Shift in Equilibrium due to the new edge (3,1). The values shown along edge $(i,j)$ is $(s_{i,j},V_{i,j})$.
$u_i$ denote total payoff of advertiser $i$.} 
\label{shift1}
\end{figure}

We also note several other observations such as 
\begin{itemize}
\item  introducing an edge may or may not shift the {\em BME} ( See Figures \ref{noshift},\ref{shift1}) 

\item  introducing an edge can shift the {\em BME} to one yielding less revenue to the auctioneer , as well as, to one with less 
efficiency i.e. social welfare ( See  Figure \ref{exfig02}, \ref{shift1})

\item  introducing new edges can shift the {\em BME} to one yielding more revenue to 
the auctioneer (See Figures \ref{shift1}, 
\ref{twobme0}, \ref{twobme1}, \ref{twobme2}) 

\item unlike in Figure \ref{shift1}, an extension of a {\em BMG} may not have a {\em BME} where 
for all nodes the degree is the same as in the {\em BME} 
for the base {\em BMG}. See Figures \ref{exfig02}, \ref{twobme1}, \ref{twobme2}. \\
\end{itemize}
 
\textbf{It is instructive to note that all the above observations (\textit{including Theorem \ref{acdilm}})
continue to hold under the solution concept of $\epsilon$-NE.}

\clearpage

\section{Auctioneer Controlled Broad Match(AcBM)}
\label{acbm}

In this section we study the second broad match scenario i.e. {\em AcBM} as described  
in section \ref{twosc}. First, we need a few definitions. 

\begin{defn}
{\bf Excess Budget:} Let $G=(\mathcal{N},\mathcal{M},\mathcal{E}$) be a {\em BMG} and 
$G^{'}=(\mathcal{N},\mathcal{M},\mathcal{E}^{'}$) be an extension (broad-match) of $G$. 
For an $i \in \mathcal{N}$, let $D_i$ denote the budget of $i$ unspent in $G$ and let 
$s_i = \max_{ j \in \mathcal{E}} s_{i,j}$. We say that $i$ has an excess budget in $G$ iff
$s_i \leq D_i$. 
\end{defn}
Intuitively, what we mean by an advertiser to have an excess budget is that she has 
enough amount currently left unspent so that she can participate in at least one query of any of the 
keywords she is currently bidding on, if given a chance, irrespective of the valuations of 
her competitors. 

\begin{defn}
{\bf Revenue Improving Broad-Match:} An extension $G^{'}=(\mathcal{N},\mathcal{M},\mathcal{E}^{'})$ 
of {\em BMG} $G=(\mathcal{N},\mathcal{M},\mathcal{E})$ is called revenue improving if there exist 
an allocation of the excess budgets along new edges (i.e. edges in $\mathcal{E}^{'} \setminus \mathcal{E} $)   
so that the sum of total budget spent by all the advertisers is more in $G^{'}$ than in $G$, as well as,
there is a strongly polynomial time algorithm to find such an allocation of the excess budgets.  
\end{defn}

Now, for a keyword $j$, let $I(j,l)$ denote the set of advertisers having sufficient budget to participate 
in the $l$th query of this keyword. The following lemma can be obtained in a way similar to the
Lemma \ref{par1} and again utilizing the partition structure per this lemma, we obtain Observation \ref{obrev}
that as long as the quality of broad-match is {\it good}
in some sense, the auctioneer can guarantee a better revenue for himself by suitably exploiting 
the extension.  

\begin{lm}
\label{par2}
Given the budget splitting of all the advertisers for the {\em BMG} $G=(\mathcal{N}, \mathcal{M}, \mathcal{E})$ 
(i.e. the current budget splitting ), that is the values $B_{i,j}$'s for all $(i,j) \in \mathcal{E}$,   
there exist non-negative integers $\Lambda_j$, $z_{j,0}, z_{j,1}, z_{j,2}, \dots, z_{j,\Lambda_j}$, and sets  
$I_{j,1}, I_{j,2}, \dots, I_{j,\Lambda_j}$, $I_j$ 
for all $j \in \mathcal{M}\}$ such that   
$\Lambda_j = O(N)$ and for all $0 \leq \lambda \leq \Lambda_j-1$, 
$c(i,j,l) = C(i,j,\lambda),  I(j,l) = I_{j, \lambda+1}  ~~ \forall z_{j,\lambda} < l \leq z_{j,\lambda+1}$    
where $C(i,j,\lambda) := c(i,j,z_{j,\lambda}+1),  I_{j, \lambda+1}:= I(j,z_{j,\lambda}+1)$ 
and
$ I_j=\{ i \in \mathcal{N}: (i,j) \in \mathcal{E}, \textrm{i  has an excess budget in $G$}  \}$. 
Moreover, all these $\Lambda, z, C, I $ values can be computed together in time polynomial in $N$, $M$ and $K$.   
\end{lm} 

\begin{obs}
\label{obrev}
Let $G^{'}=(\mathcal{N},\mathcal{M},\mathcal{E}^{'})$ be an extension
of {\em BMG} $G=(\mathcal{N},\mathcal{M},\mathcal{E})$,  $I= \cup_{j \in \mathcal{M}} I_j$, $
J_i = \left\{ j \in \mathcal{M}: (i,j) \in \mathcal{E}^{'} \setminus \mathcal{E}\right\}$ 
for  $ i \in I$, 
$\Gamma_j = \left\{ i \in \mathcal{N} : (i,j) \in \mathcal{E}^{'}\right\}$, and 
$\Phi = \left\{ j \in \mathcal{M}: I_{j,\Lambda_j}=\phi \right\}$. This extension is revenue improving if 
there is an $i \in I$ having one or more of the following properties/conditions: 
\begin{itemize}
\item[a)] $\exists j \in J_i \setminus \Phi$ such that $I_{j,\Lambda_j} = I_j$ and 
$| \left\{ m \in I_{j,\Lambda_j}: s_{m,j} > s_{i,j}\right\}| < K$.
\item[b)] $\exists j \in J_i \cap \Phi$ such that 
$| \left\{ l \in I \cap \Gamma_j: s_{l,j} = s_{i,j}\right\} \cup \left\{ l \in \Gamma_j: s_{l,j} < s_{i,j}\right\}| > 1$.
\item[c)] $\exists j \in J_i \setminus \Phi$ such that $I_{j,\Lambda_j} \neq I_j$ and 
$s_{i,j} > \max_{ l \in I_{j,\Lambda_j} \setminus I_j } s_{l,j}$ .  
\end{itemize}
\end{obs}
Intuitively, the properties $a)$ and $c)$ says that there is an advertiser $i$ with excess budget in $G$ and 
a keyword $j$ such that $i$ has a good enough valuation for $j$ in $G^{'}$ so as 
to obtain a slot, and moreover if $i$ is brought in for the last query, all
advertisers already participating in that query have still 
enough budget to participate in that query.  The condition $b)$ says that there is a keyword $j$ with unsold queries 
in $G$ and we can sell these queries in $G^{'}$ to at least two
advertisers\footnote[13]{Note that just selling to one advertiser does not generate any money in GSP. 
In practice, however Google/Yahoo! charges a minimum amount i.e. a reserve price for the last slot.
Nevertheless, all the results we present remains unchanged by introducing reserve prices for the last slot.
In that case, the condition $b)$ will change to $\exists j \in J_i \cap \Phi$.}.
Thus, these conditions allow the auctioneer to bring additional advertiser(s) for the keyword $j$ to $j$'s last query or equivalently 
the very first query in the last partition ($(z_{j,\Lambda_j-1}+1)$th query) 
so that revenue extracted from this particular query is improved without 
changing the revenue generated from any other query. 
Therefore, in terms of the existence of a revenue improving 
extension we have the above observation.    
Nevertheless, in general the auctioneer can do much more than the above trivial way.
In particular, there might be several $i$ and $j$'s satisfying 
the properties in Observation \ref{obrev} and he could exploit this profitably. Recall that he has a finer control over 
which query to start with and whatever part of excess budget he can 
spend along edges in $\mathcal{E}^{'} \setminus \mathcal{E}$. 
So his task is to choose splitting of excess budgets along new edges as well as to decide a starting query number.
In general, there could be $O(N)$ advertisers with excess budgets and there could be $O(MN)$ new edges, finding the 
best splitting is a variant of Integer Knapsack problem again and thus computationally hard. 
Nevertheless, it is clearly possible to design strongly polynomial time sub-optimal algorithm that does significantly 
better than the trivial improvement possible by increasing competition for the last query. The problem with participating 
starting a query in other partitions is that it may change the partition structure in a way that is not 
revenue improving, but since there are only $O(N)$ such partitions for each keyword, we could
check for each partition whether starting with its first query is revenue improving or not. Indeed, 
it is easy to think of a strongly polynomial time algorithm that finds 
excess budget splittings which improves revenue, one that moves from partition to partition, taking starting query as 
the first query in that partition and then doing a binary search for appropriate budget (as high as possible) to be allocated and
tracks which one of these possibilities lead to the highest revenue. Note that since we are doing a binary search on budget
(and equivalently on the number of queries to participate in), we 
are still in strongly polynomial time regime. Finally, it should be interesting to search for efficient algorithms 
generating better revenue and in particular a FPTAS, possibly by efficiently searching for which query to start with
along with efficiently searching for how many queries to participate for. \\
 
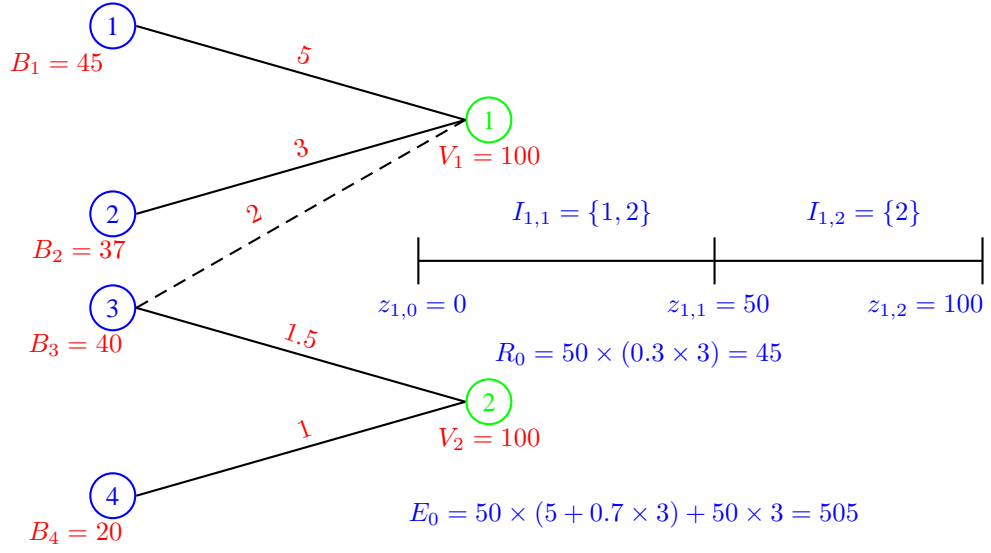
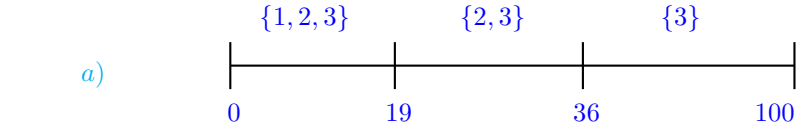
\begin{figure}[h] 
\begin{center} 
\psset{unit=1.25cm}
\begin{pspicture}(0, -10.8)(4, 6)
\psline(-0.75,0)(2.75,1) 
\rput{20}(1,0.7){ \small \color{red} $1$}

\pscircle[linecolor=blue](-1,0){0.25}
\rput(-1,0){\small \color{blue} 4}
\rput(-1.4,-0.4){\small \color{red} $B_4=20$}

\pscircle[linecolor=green](3,1){0.25} 
\rput(3,1){\small \color{green} 2}
\rput(3,0.6){\small \color{red} $V_2=100$}

\psline(-0.75,2)(2.75,1) 
\rput{-20}(1,1.7){\small \color{red} $1.5$}

\pscircle[linecolor=blue](-1,2){0.25} 
\rput(-1,2){\small \color{blue} 3}
\rput(-1.4,1.6){\small \color{red} $B_3=40$}

\psline(-0.75,3)(2.75,4) 
\rput{20}(1,3.7){\small \color{red} $3$}

\pscircle[linecolor=blue](-1,3){0.25} 
\rput(-1,3){\small \color{blue} 2}
\rput(-1.4,2.6){ \small \color{red} $B_2=37$}

\pscircle[linecolor=green](3,4){0.25} 
\rput(3,4){\small \color{green} 1}
\rput(3,3.6){\small \color{red} $V_1=100$}

\psline(-0.75,5)(2.75,4) 
\rput{-20}(1,4.7){ \small \color{red} $5$}

\pscircle[linecolor=blue](-1,5){0.25} 
\rput(-1,5){\small \color{blue} 1}
\rput(-1.6,4.6){\small \color{red} $B_1 = 45$}


\psline[linestyle=dashed](-0.75,2)(2.75,4) 
\rput{26}(0.5,3){\small \color{red} $2$}


\psline(2.25,2.5)(8.25,2.5) 
\psline(2.25,2.75)(2.25,2.25)
\psline(5.4,2.75)(5.4,2.25) 
\rput(2.25,2){ \small \color{blue} $z_{1,0}=0$}
\rput(5.4,2){ \small \color{blue} $z_{1,1}=50$}
\rput(3.95,3){ \small \color{blue} $I_{1,1}=\{1,2\}$}
\rput(4.55,1.5){ \small \color{blue} $R_0=50 \times (0.3 \times 3) = 45$}
\rput(4.5, -0.2){ \small \color{blue} $E_0=50 \times (5+0.7 \times 3)+ 50 \times 3=505$}
\psline(8.25,2.75)(8.25,2.25) 
\rput(7.6,2){ \small \color{blue} $z_{1,2}=100$}
\rput(6.95,3){ \small \color{blue} $I_{1,2}=\{2\}$}

\psline(0.25,-2)(6.25,-2) 
\psline(0.25,-1.75)(0.25,-2.25)
\psline(2,-1.75)(2,-2.25) 
\psline(4,-1.75)(4,-2.25) 

\rput(-1.25,-2.1){ \small \color{cyan} $a)$}

\rput(0.25,-2.5){ \small \color{blue} $0$}
\rput(2,-2.5){ \small \color{blue} $19$}
\rput(1,-1.5){ \small \color{blue} $\{1,2,3\}$}

\rput(4,-2.5){ \small \color{blue} $36$}
\rput(3,-1.5){ \small \color{blue} $\{2,3\}$}

\psline(6.25,-1.75)(6.25,-2.25) 
\rput(6,-2.5){ \small \color{blue} $100$}
\rput(5,-1.5){ \small \color{blue} $\{3\}$}

\rput(2,-3){ \small \color{blue} $R_a=19 \times (0.3 \times 3 + 2 \times 1.4 \times 2) + 17 \times (0.3 \times 2) = 80.5 > R_0$}
\rput(2, -3.5){ \small \color{blue} $E_a=19 \times (5+0.7 \times 3)+ 17 \times (3 + 0.7 \times 2) + 64 \times 2=337.7 < E_0$}

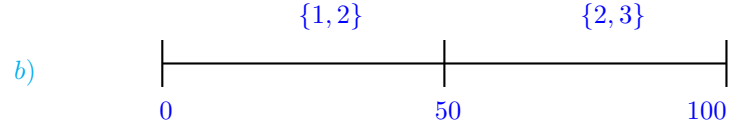
\psline(0.25,-5)(6.25,-5) 
\psline(0.25,-4.75)(0.25,-5.25)
\psline(3.25,-4.75)(3.25,-5.25) 

\rput(-1.25,-5.1){ \small \color{cyan} $b)$}

\rput(0.25,-5.5){ \small \color{blue} $0$}
\rput(2,-4.5){ \small \color{blue} $\{1,2\}$}

\rput(3.25,-5.5){ \small \color{blue} $50$}

\psline(6.25,-4.75)(6.25,-5.25) 
\rput(6,-5.5){ \small \color{blue} $100$}
\rput(5,-4.5){ \small \color{blue} $\{2,3\}$}

\rput(2,-5.9){ \small \color{blue} $R_b=50 \times (0.3 \times 3) + 50 \times (0.3 \times 2) = 75 > R_0$}
\rput(2, -6.4){ \small \color{blue} $E_b=50 \times (5+0.7 \times 3)+ 50 \times (3 + 0.7 \times 2)= 575 > E_0$}

\rput(2, -7.2){ \small \color{blue} \framebox{$R_0 < R_b < R_a$  but  $E_a < E_0 < E_b$}}

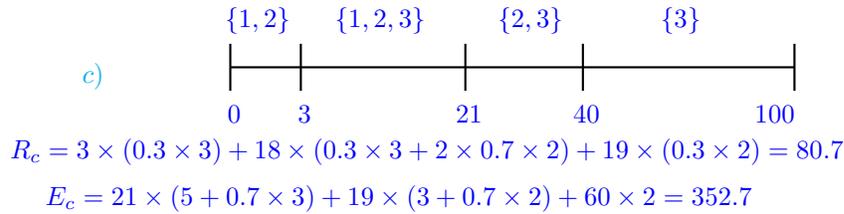
\psline(0.25,-9)(6.25,-9) 
\psline(0.25,-8.75)(0.25,-9.25)
\psline(4,-8.75)(4,-9.25) 
\psline(1,-8.75)(1,-9.25) 
\psline(2.75,-8.75)(2.75,-9.25) 

\rput(-1.25,-9.1){ \small \color{cyan} $c)$}

\rput(0.25,-9.5){ \small \color{blue} $0$}
\rput(0.5,-8.5){ \small \color{blue} $\{1,2\}$}
\rput(1.8,-8.5){ \small \color{blue} $\{1,2,3\}$}
\rput(3.4,-8.5){ \small \color{blue} $\{2,3\}$}
\rput(4,-9.5){ \small \color{blue} $40$}
\rput(1,-9.5){ \small \color{blue} $3$}
\rput(2.75,-9.5){ \small \color{blue} $21$}

\psline(6.25,-8.75)(6.25,-9.25) 
\rput(6,-9.5){ \small \color{blue} $100$}
\rput(5,-8.5){ \small \color{blue} $\{3\}$}

\rput(2.3,-9.9){ \small \color{blue} $R_c=3 \times (0.3 \times 3)+  18 \times (0.3 \times 3 + 2 \times 0.7 \times 2) + 19 \times (0.3 \times 2) = 80.7$}
\rput(2, -10.4){ \small \color{blue} $E_c=21 \times (5+0.7 \times 3)+ 19 \times (3 + 0.7 \times 2) + 60 \times 2= 352.7$}

\rput(3.5, -11.2){ \small \color{blue} \framebox{$R_c >  R_a$}}

\end{pspicture}
\end{center} 
\caption{Tradeoff in revenue and social welfare, the base {\em BMG} does not
include edge $(3,1)$, its extension does. The revenue and efficiency value is just for keyword $1$, as 
these values do not change for keyword $2$ even in the extension. In a) and b) advertiser $3$ is brought in for keyword $1$ 
starting with first query in partition $1$ and $2$ respectively. Note that the choice of partition with better revenue is the one 
where efficiency decreases. $c)$ shows that just starting with the first query of a partition may not lead to optimal revenue, for example 
if $3$ is started with $4$th query then the revenue improvement is even better than $a)$ or $b)$. } 
\label{acbmexfig1}
\end{figure}

Now, given that auctioneer's goal is primarily 
to improve revenue, we should also analyze what happens to the efficiency (i.e. social welfare), if the auctioneer 
implements such revenue improving broad-match. First, it is clear that if auctioneer's goal were to improve 
social welfare instead of revenue, he could certainly do so in a way similar to the one discussed above for revenue improvement.
Moreover, revenue and social welfare could infact be  improved together if one of the conditions 
in Observation \ref{obrev} holds
and the proof is same 
as that for Observation \ref{obrev} by bringing in appropriate advertiser(s) in the last query of an appropriate keyword. 
However, if auctioneer deviates from this trivial way of improving revenue, which infact he will do if his goal is to 
maximize revenue, there can often be a tradeoff with social welfare. For an explicit 
example of this tradeoff please refer to Figure \ref{acbmexfig1}. 
Furthermore,   
{\it even when the conditions in  Observation \ref{obrev} are NOT satisfied, there might still 
be a possibility of revenue improvement} (ref. Figure \ref{exfig02}). \\

\clearpage

\section{Future Directions}
\label{concl}
We have initiated a study  
of broad-match, an interesting aspect of sponsored search advertising,
and as common to papers that initiate a new direction of study, this paper leaves out several 
important open problems that deserve theoretical investigation and analysis. 
We discuss some of these interesting problems in the following. 

\paragraph*{Budget Splitting Games(BSG):} Abstracting the settings in Broad-Match Game can provide us with a 
rich class of mutli-player games having compact representations. It should be interesting 
to study these games and to consider the budget splitting/allocation scenarios beyond the currently prevailing 
models for sponsored search advertising, as well as, other interesting applications. 
Herebelow we provide an abstraction.

 An instance of \emph{BSG} is given by
 $(\mathcal{N},\mathcal{M},\mathcal{E}, B,V,\mathcal{O})$. 
$\mathcal{N}$ is the set of players, $\mathcal{M}$ is set of distinct type of \emph{indivisible} items 
and $(\mathcal{N},\mathcal{M},\mathcal{E})$ is a bi-partite graph between the players and items
wherein an $(i,j) \in \mathcal{E} $ iff the player $i\in \mathcal{N}$ is interested in buying the 
item $j\in \mathcal{M}$ (i.e. $i$ have a positive valuation for $j$). $B=\{B_i\}_{i \in \mathcal{N}}$ 
is the budget vector, $B_i$ being the total budget of player $i$. 
$V=\{V_j\}_{j \in \mathcal{M}}$ 
is the volume vector, $V_j$ being the total number of units of the item $j$ for sale. 
Let $|\mathcal{N}|=N$ and $|\mathcal{M}|=M$. 
$\mathcal{O}$ is an oracle that takes as input $(i,j) \in \mathcal{E}$ and a set of budget values 
$\{B_{n,j}\}_{n \in \mathcal{N}-\{i\}}$ and output the set of values 
$\Lambda_j = O(N)$, $z_{j,0}=0, z_{j,1}, z_{j,2}, \dots, z_{j,\Lambda_j}=V_j$, and 
$C(i,j,\lambda), U(i,j,\lambda)$ for all $0 \leq \lambda \leq \Lambda_j-1$, to be 
interpreted as follows: given that for all $n \in \mathcal{N}-\{i\}$, the player $n$ spends/allocates an amount 
$B_{n,j}$ on the item $j$, $C(i,j,\lambda)$ and $U(i,j,\lambda)$ respectively are the cost and the utility 
for each unit of the item if the player $i$ decides to buy 
$ z_{j,\lambda} < V_{i,j} \leq  z_{j,\lambda+1}$ units of the item $j$. 
Each player has access to the oracle $\mathcal{O}$. The strategy of a player $i$ is to split her budget 
across various items that is to 
choose the values $\{B_{i,j}\}_{j \in \mathcal{M}:(i,j) \in \mathcal{E}}$ 
such that $\sum_{j \in \mathcal{M}:(i,j) \in \mathcal{E}}B_{i,j}\leq B_i$ and correspondignly 
the number of units of the various items $\{V_{i,j}\}_{j \in \mathcal{M}:(i,j) \in \mathcal{E}}$ with $V_{i,j}\leq V_j$.
The payoff of a player is the total utility it derives across all the items from their units she bought 
as per her chosen strategy. 

Note that this abstraction nicely captures the Broad-Match Game studied in the present paper wherein 
an item is a keyword and a query of a keyword corresponds to a unit. The oracle $\mathcal{O}$
corresponds to the Lemma \ref{par1}.  Further, by varying the properties satisfied 
by the values $C(i,j,\lambda)$, $U(i,j,\lambda)$'s we may obtain other interesting scenarios.

\paragraph*{Complexity of Auctioneer's Dilemma:} 
The Theorem \ref{acdilm} leaves the auctioneer in a dilemma about whether he should 
broad-match or not. If he could somehow predict which choice of broad match
lead to a revenue improvement for him and which not, he could potentially 
choose the ones leading to a revenue improvement. 
This brings 
forth one of the big questions left open in this paper, 
that is of efficiently computing a  \textit{BME} / \textit{$\epsilon$-NE}, if one exists, 
given a choice of \textit{broad match},
as well as, to efficiently decide whether a given broad match is 
revenue improving or not in the {\em AdBM} scenario. 

\paragraph*{Existence of pure NE/$\epsilon$-NE/BME in Broad-Match Game:}
Despite some effort we have not been able to show that a Broad-Match Game always 
admits pure NE or even $\epsilon$-NE or BME. On the other hand we have also not been 
able to construct examples where they do not exist. 
The problem arises from the fact that the payoff funtions are discontinuous 
and not even quasi-concave. Therefore, the existing techniques or their 
simple extensions do not seem to work. 
This calls for developing new techniques for proving the existence of pure 
Nash equilibria and the complexity of deciding the existence in general, and  
for the Broad-Match Game and BSGs in particular. 

\paragraph*{Quantifying the Effect of Broad-Match:} 
As per various observations including Theorem \ref{acdilm} obtained in the Section \ref{bmnbm} 
we know that the the revenue of the autioneer 
and the social welfare could very well degrade by incorporating broad-match and it would be 
interesting to characterize the extent of such degradation. 

\paragraph*{Total Click-Through-Rates as Payoff:} 
It would be interesting to analyze the effect of broad-match in a 
scenario where the advertisers are interested in maximizing their 
total Click-Through-Rate across various keywords rather than their 
total utilities. Note that the budget splitting game arising 
in this sceneario indeed fits in our abtract model of \emph{BSG} discussed 
earlier in this section. With the intuition developed 
from this paper, we believe that similar conclusions hold true even 
in the case of total Click-Through-Rates as payoffs.

\paragraph*{BME vs $\epsilon$-NE:}
As we argued in the paper, both the \emph{BME} and \emph{$\epsilon$-NE} are 
reasonable solution concepts for the Broad-Match Game. 
Further, we constructed examples where the both solution concepts coincide.  
Nevertheless, it would be nice to study these two concept comparatively 
at a finer level. For example, \emph{BME} may not provide good approximation guarantee 
to the advertisers but it demands less shophistication level from them than that required 
for good approximation guarantee. 

\paragraph*{Network Level Competition via  Broad-Match Game:}  
The framework provided in the present paper is easily applicable to 
the scenario where advertisers are 
trying to optimally split budgets across various keywords 
\textit{coming from several competing search engines.} 
How does the 
revenue of one search engine gets effected by the fact that another 
competeting search engine offers broad-match?  
For example consider the BMG in the Figure \ref{shift1}, 
without the dashed edge $(3,1)$, and 
suppose that the keyword $1$ is coming from the search engine $1$ and 
the keyword $2$ is coming from the serach engine $2$. 
Now if the search engine $1$ does broad-match and introduces the new edge $(3,1)$ then 
we can see that the revenue of the search engine $1$ increases whereas that of search engine $2$ decreases. 
It should be interesting to further explore this direction.

\paragraph*{Formalizing a Notion of ``Good'' Broad-Match:}
In the analysis we have presented,  
we have not considered costs incurred by the auctioneer in the short run
due to uncertainty about the extended part of the {\em BMG}, although 
we can expect that as long as the quality of broad match being performed 
is good, such costs should be minimal. Nevertheless,
formalizing a notion of {\it good} broad match should be interesting.
Features such as the improvement in the relevance scores and
conversion rates should be an essential ingredient of a {\it good} broad match. 
Further, the conditions noted in the Observation \ref{obrev} also provides 
some sense of what should be the features of a good broad-match 
from the viewpoint of the auctioneer.  

\paragraph*{Acknowledgements:} 
We thank Gunes Ercal, Himawan Gunadhi, Adam Meyerson and Behnam Rezaei for discussions and anonymous
reviewers of an earlier version for valuable comments and suggestions. 
SKS thanks Zo\"e Abrams, Arpita Ghosh, Gagan Goel, Jason Hartline, Aranyak Mehta,  Hamid Nazerzadeh, and Zoya Svitkina for 
useful discussions during WINE 2007.


\begin{thebibliography}{100}

\bibitem{AGM06} G. Aggarwal, A. Goel, R. Motwani, Truthful Auctions for Pricing Search Keywords, EC 2006.
\bibitem{BCI+07} C. Borgs, J. Chayes, N. Immorlica, K. Jain,  O. Etesami, and M. Mahdian, 
Dynamics of bid optimization in online advertisement auctions, WWW 2007. 
\bibitem{CDE+07} M. Cary, A. Das, B. Edelman, I. Giotis, K. Heimerl, A. R. Karlin, C. Mathieu, and M. Schwarz,
Greedy bidding strategies for keyword auctions, EC 2007. 
\bibitem{EO07} B. Edelman and M. Ostrovsky, Strategic bidder behavior in sponsored search auctions, Decis. Support Syst.,
43(1):192-198, 2007.  
\bibitem{EOS05} B. Edelman, M. Ostrovsky,  M. Schwarz,
Internet Advertising and the Generalized Second Price Auction: Selling Billions of Dollars Worth of Keywords, 
American Economic Review, 97(1):242-259, 2007.
\bibitem{FMPS07} J. Feldman, S. Muthukrishnan, M. Pal, and C. Stein, Budget optimization in search-based advertising auctions,
EC 2007. 
\bibitem{GP07} R. Gonen and E. Palkov, An Incentive-Compatible Multi-Armed Bandit Mechanism, PODC 2007. 
\bibitem{IK75} O. H. Ibarra and C. E. Kim, 
Fast Approximation Algorithms for the Knapsack and Sum of Subset Problems, JACM , Vol. 22, pp:463 - 468 , 1975. 
\bibitem{Lahaie06} S. Lahaie, An Analysis of Alternative Slot Auction Designs for Sponsored Search, EC 2006.
\bibitem{Lawler79} E. L. Lawler, Fast Approximation Algorithms for Knapsack Problems
Mathematics of Operations Research, Vol. 4, pp: 339-356, 1979. 
\bibitem{LP07} S.  Lahaie, and D. Pennock, Revenue Analysis of a Family of Ranking Rules for Keyword Auctions, EC 2007.
\bibitem{MSVV05} A. Mehta, A.   Saberi, U.  Vazirani, and V.   Vazirani, AdWords and generalized on-line matching, FOCS 2005.
\bibitem{MPS07} S. Muthukrishnan, M. P{\'a}l, and Z. Svitkina, 
Stochastic Models for Budget Optimization in Search-Based Advertising, WINE 2007. 
\bibitem{papa07} C. H. Papadimitriou, The Complexity of Finding Nash Equilibria, 
Chapter 2, Algorithmic Game Theory, Cambridge University Press, 2007.
\bibitem{PR05} C. H. Papadimitriou, and T. Roughgarden, Computing Equilibria in Multi-Player Games, SODA 2005. 
\bibitem{PSbook}  C. H. Papadimitriou, and K. Steiglitz, 
\emph{Combinatorial Optimization: Algorithms and Complexity}, Dover Publications 1998. 
\bibitem{Roughgarden05} T. Roughgarden, Selfish Routing and the Price of Anarchy, MIT Press, 2005. 
\bibitem{RW06} P. Rusmevichientong, and D. P. Williamson, 
An adaptive algorithm for selecting profitable keywords for search-based advertising services, EC 2006. 
\bibitem{SRBR08} S. K. Singh, V. P. Roychowdhury,  M. Bradonji\'c, and B. A. Rezaei,
Exploration via design and the cost of uncertainty in keyword auctions, Submitted. 
\bibitem{SRGR07} S. K. Singh, V. P. Roychowdhury, H. Gunadhi, and B. A. Rezaei,
Capacity Constraints and the Inevitability of Mediators in Adword Auctions, WINE 2007. 
\bibitem{SRGR08} S. K. Singh, V. P. Roychowdhury, H. Gunadhi, and B. A. Rezaei,
Diversification in the Internet Economy:The Role of For-Profit Mediators,
Submitted. 
\bibitem{Varian06} H. Varian, Position Auctions, International Journal of Industrial Organization, 25(6):1163-1178, 2007.
\bibitem{vazapproxbook} V. V. Vazirani, \emph{Approximation Algorithms}, Springer 2001. 
\bibitem{WVLL07} J. Wortman, Y. Vorobeychik, L. Li, and J. Langford,
Maintaining equilibria during exploration in sponsored search auctions, WINE 2007.

\end{thebibliography}
\end{document}